\documentclass[aps,prd,onecolumn,eqsecnum,amsmath,nofootinbib,preprintnumbers]{revtex4}%
\usepackage{color,graphicx,float,subfigure,xcolor}
\usepackage{amsfonts,amssymb,theorem,mathrsfs,times}
\usepackage{bm}
\usepackage{multirow}
\usepackage{mathtools}
\usepackage{amsfonts,amssymb,theorem,mathrsfs}
\usepackage{dsfont}
\usepackage{setspace}
\usepackage{amsmath} 
\usepackage{amssymb}
\usepackage{marvosym}
\usepackage{fontawesome}
\usepackage{pifont}
\usepackage[colorlinks,linkcolor=blue,anchorcolor=blue,citecolor=blue]{hyperref}   
\usepackage{ulem}   
\usepackage[english]{babel}
\usepackage{physics}

{\theorembodyfont{\upshape}
	}
{\theorembodyfont{\upshape}
	}
{\theorembodyfont{\upshape}
	}
{\theorembodyfont{\upshape}
	}
{\theorembodyfont{\upshape}
	}
{\theorembodyfont{\upshape}
	}
\addtocounter{MaxMatrixCols}{10}
\newcommand{\dalm}{\kern1pt\vbox{\hrule height 0.9pt\hbox{\vrule width 0.9pt\hskip 2.5pt\vbox{\vskip 5.5pt}\hskip 3pt\vrule width 0.3pt}\hrule height 0.3pt}\kern1pt}

\begin{document}
\thispagestyle{empty}
\title{Quasinormal modes of Schwarzschild-de Sitter black holes in semi-open systems}
	
%
\author{Liang-Bi Wu$^{a}$\footnote{e-mail address: liangbi@mail.ustc.edu.cn}}

\author{Libo Xie$^{a\, ,b\, ,c}$\footnote{e-mail address: xielibo23@mails.ucas.ac.cn (corresponding author)}}

\author{Li-Ming Cao$^{d}$\footnote{e-mail address: caolm@ustc.edu.cn}}

\author{Ming-Fei Ji$^d$\footnote{e-mail address: jimingfei@mail.ustc.edu.cn}}

\author{Yu-Sen Zhou$^d$\footnote{e-mail address: zhou\_ys@mail.ustc.edu.cn}}	 


\affiliation{${}^a$School of Fundamental Physics and Mathematical Sciences, Hangzhou Institute for Advanced Study, UCAS, Hangzhou 310024, China}

\affiliation{${}^b$CAS Key Laboratory of Theoretical Physics, Institute of Theoretical Physics, Chinese Academy of Sciences, Beijing 100190, China}

\affiliation{${}^c$University of Chinese Academy of Sciences, Beijing 100049, China}

\affiliation{${}^d$Interdisciplinary Center for Theoretical Study and Department of Modern Physics, University of Science and Technology of China, Hefei, Anhui 230026, China}

\date{\today}
	
\begin{abstract}
We study perturbations of Schwarzschild-de Sitter black holes in semi-open systems by using the Heun functions. For the semi-open system, a partially reflective wall is added around the event horizon. Three aspects of this model are investigated, namely the quasinormal mode (QNM) spectra, the greybody factor (GF), and the exceptional point (EP). For the QNM aspect, we identify three distinct behaviors as the frequency-independent reflectivity $\mathcal{K}$ increasing. The first-type modes approach the real axis and form long-lived quasi-bound states. The second-type modes move toward but do not reach the real axis and retain a finite decay rate. The third-type modes eventually lie on the imaginary axis becoming purely decaying modes. For the GF aspect, GFs exhibit strong oscillations controlled by the distance between the potential and the reflective wall with a real constant reflectivity. In contrast, a Boltzmann-type reflectivity produces only small corrections. Finally, by promoting $\mathcal{K}$ to a complex parameter, the modified boundary conditions give rise to a second-order EP. Parameterizing the vicinity of such EP, we observe the mode exchange phenomenon, and the deviation of spectra scale with the square root of the deviation of the parameter, as predicted by a Puiseux series expansion.

\end{abstract}

\maketitle
	
\section{Introduction}\label{Introduction}
When characterizing the intrinsic oscillations of perturbed black holes, quasinormal modes (QNMs) appear as a discrete set of complex frequencies. The real part corresponds to the oscillation frequency, and the imaginary part reflects the damping rate. Serving as unique spectral fingerprints of black holes, QNMs provide a powerful probe to test the Kerr hypothesis and to investigate gravity in the strong-field regime. This field of study is referred  to as black hole spectroscopy~\cite{Kokkotas:1999bd,Berti:2009kk,Konoplya:2011qq,Berti:2025hly}.

The QNM spectra of black holes exhibit the instability~\cite{Konoplya:2022pbc,Jaramillo:2020tuu,Shen:2025yiy} due to the non-Hermitian (NH) of black hole systems~\cite{Ashida:2020dkc}, and this phenomenon has been extensively studied. That is to say, these spectra respond with significant shifts in the complex plane to the slightest external perturbations. Initial studies of QNM spectrum instability for the black holes were provided by Nollert and Price~\cite{Nollert:1996rf,Nollert:1998ys}. Methods for investigating spectrum instability associated with QNMs can be classified into two primary categories. Pseudospectrum analysis~\cite{trefethen2020spectra} is a reliable tool to investigate the instability of the QNM spectrum~\cite{Boyanov:2024fgc, Jaramillo:2020tuu,Destounis:2023ruj,Jaramillo:2021tmt}. This method employs visual techniques with the goal of elucidating spectrum instability by analyzing the characteristic properties of non-self-adjoint operators in dissipative systems. Within the gravity theory, pseudospectra provide qualitative insights into spectrum instability across different spacetimes~\cite{Jaramillo:2020tuu,Destounis:2021lum,Cao:2024oud,Cao:2024sot,Arean:2024afl,Garcia-Farina:2024pdd,Arean:2023ejh,Boyanov:2023qqf,Cownden:2023dam,Carballo:2025ajx,Sarkar:2023rhp,Destounis:2023nmb,Luo:2024dxl,Warnick:2024usx,Chen:2024mon,Boyanov:2022ark,Siqueira:2025lww,Besson:2024adi,dePaula:2025fqt,Cai:2025irl,Cao:2025qws}. Transient dynamics associated with pseudospectra are investigated in~\cite{Carballo:2024kbk,Jaramillo:2022kuv,Chen:2024mon,Carballo:2025ajx,Besson:2025ghu}.

However, the information provided by pseudospectra regarding spectrum stability is insufficient. This can be understood from its definition: for a given perturbation magnitude under a specific norm, the maximal spectrum shift precisely corresponds to the boundary of the pseudospectrum. While valuable, its inability to identify the precise positions of spectrum shifts greatly limits its application. Accordingly, one may consider the so-called structured pseudospectrum~\cite{trefethen2020spectra}. In recent work, attention has increasingly shifted to understanding how the spectrum responds to perturbations of a given form, which are commonly added to the effective potential. Therefore, one modifies the effective potential to study the QNM spectrum instability~\cite{Qian:2020cnz,Daghigh:2020jyk,Liu:2021aqh,Li:2024npg,Qian:2024iaq,Xie:2025jbr,Berti:2022xfj,Cheung:2021bol,Yang:2024vor,Courty:2023rxk,Cardoso:2024mrw,Ianniccari:2024ysv,MalatoCorrea:2025iuc,Shen:2025nsl,Wang:2025mxe,Hu:2025efp}. In these references, more recently, black hole spectrum instability triggered deterministic and random metric perturbations is studied in~\cite{Shen:2025nsl}. For a more realistic physical model, namely Einstein-Maxwell-scalar (EMS) model, it is found that the stable and the unstable modes will coexist~\cite{Wang:2025mxe}. For dynamical Chern-Simons gravity, three distinctive phenomena absent in general relativity are found in terms of the QNM spectrum instability~\cite{Hu:2025efp}. When the small extra bump becomes negative enough, spacetime will even become unstable~\cite{Mai:2025cva}. The above studies are all considered within the standard QNM boundary conditions, where the ingoing boundary condition is at the event horizon and outgoing boundary condition is at infinity or the cosmological horizon.

Considering that black holes in reality are not perfect, that is, they are not completely ``black'', which will be replaced by the exotic compact objects (ECOs). In other words, there exists a wall with the reflectivity $\mathcal{K}(\omega)$ at $x_0$ in the tortoise coordinate. For these systems of ECOs, mathematically speaking, the boundary conditions around the event horizon will be modified into $\Psi\sim \mathrm{e}^{-\mathrm{i}\omega(x-x_0)}+\mathcal{K}(\omega)\mathrm{e}^{\mathrm{i}\omega (x-x_0)}$ at the frequency-domain. One of the resulting features of the ringdown for the ECOs is the gravitational wave (GW) echoes~\cite{Mark:2017dnq,Cardoso:2019apo,Bueno:2017hyj,Cardoso:2016rao,Cao:2025qxt,Zhang:2025ygb,Yang:2025hqk}. The reflectivity plays a key role in both modulating the signal and determining the amplitude of subsequent echoes. A list of models for the reflectivity has been explicitly work out such as the wormhole model~\cite{Mark:2017dnq,Bueno:2017hyj}, the Boltzmann model~\cite{Oshita:2019sat,Wang:2019rcf}, the area quantization model~\cite{Deppe:2024fdo}, the perfect-fluid star, gravastar, membrane paradigm models~\cite{Chakraborty:2023zed}. One can find a table summarizing the above models in~\cite{Rosato:2025byu,Berti:2025hly}. It has been universally applicable that the QNM spectra of black holes are unstable against small modifications, so even small changes in boundary conditions may lead to the QNM spectrum instability~\cite{Oshita:2025ibu,Solidoro:2024yxi}. Recently, ECOs version of flea and elephant models has been studied in~\cite{Destounis:2025dck}.

According to the Green's function theory, if we can know the so-called ``in'' solution and ``up'' solution, we can calculate the (ringdown) waveforms of gravitational waves generated by any form of source in principle~\cite{Andersson:1995zk,Berti:2006wq,Zhang:2013ksa,Berti:2025hly}. If the ``in'' solution and ``up'' solution can be analytically solved, the technical difficulty of high-precision waveform construction will be greatly reduced. According to our research, there are P\"{o}shl-Teller (PT) models~\cite{Ferrari:1984zz}, BTZ black hole~\cite{Arnaudo:2025uos}, Schwarzschild-de Sitter (SdS) black hole~\cite{Arnaudo:2025uos,Arnaudo:2025kit}, Kerr-de Sitter black hole~\cite{Hatsuda:2020sbn} and $\text{Kerr-AdS}_5$ spacetime that can be analytically solved~\cite{Noda:2022zgk}. Specifically, for the SdS black hole, QNMs are able to be solved by the Heun function~\cite{Arnaudo:2025uos,Arnaudo:2025kit}. One can refer to~\cite{Chen:2025sbz} for a complete treatment of QNMs of Type-D black holes, in which the Heun function is also used. In addition, for the Schwarzschild black hole, the Coulomb wave functions are used to construct the Green function~\cite{Leaver:1986gd,Berti:2025hly}. More recently, in the Grumiller spacetime~\cite{Mi:2025fbt} and in the rotating Kalb-Ramond BTZ black hole~\cite{Xia:2025hwt}, QNMs can be solved analytically.

In this study, we focus on the perturbations of the SdS black hole. The SdS spacetime models a non-spinning, uncharged black hole in an asymptotically de Sitter universe, being a black hole solution with a positive cosmological constant $\Lambda>0$. Its astrophysical importance arises from cosmological evidence for a positive cosmological constant, meaning real astrophysical black holes reside in spacetimes where this constant, though small, modifies the large-scale asymptotic geometry. The QNM spectra of SdS black hole have been computed in~\cite{Zhou:2025xta,Cardoso:2003sw,Konoplya:2004uk,Jansen:2017oag,Cardoso:2017soq,Konoplya:2022xid} by pseudo-spectral method, Leaver method or near-extreme approximation method, respectively. One of our studies in this work is to show how the spectra of the SdS black hole are affected by reflectivity $\mathcal{K}(\omega)$ being near the event horizon.

While frequency-domain analyses reveal that small disturbances can cause substantial spectral changes, as demonstrated by studies of effective potential modifications, pseudospectrum approaches and boundary condition modifications, the ringdown waveform remains stable~\cite{Wu:2025sbq,Daghigh:2020jyk,Berti:2022xfj,Yang:2024vor,Oshita:2025ibu,Kyutoku:2022gbr}. The stability of waveform and the stability of greybody factor (GF) are inseparable~\cite{Kyutoku:2022gbr}. In other words, the black hole greybody factors are more robust observables~\cite{Oshita:2023cjz,Rosato:2024arw,Oshita:2024fzf,Ianniccari:2024ysv,Wu:2024ldo,Kyutoku:2022gbr,Konoplya:2025ixm,Wu:2025sbq,Xie:2025jbr} unlike the QNM spectra. Although QNMs and GFs are both characteristic quantities reflecting the geometry of spacetime, they are defined under different boundary conditions. Their correspondence was first established for spherically symmetric black holes in~\cite{Konoplya:2024lir}. The correspondence between QNMs and GFs has been extended to various black holes~\cite{Dubinsky:2025nxv,Konoplya:2024vuj,Bolokhov:2024otn,Skvortsova:2024msa,Malik:2024cgb}. For ultracompact horizonless objects, Refs. \cite{Rosato:2025byu,Rosato:2025lxb} give detailed discussion on the GFs. Therefore, motivated by them, we also investigate the GFs on the SdS black hole within the existence of the reflectivity. Here, we focus on two cases. One is that $\mathcal{K}(\omega)$ is a real constant independent of the frequency, while the other is that $\mathcal{K}(\omega)$ is the Boltzmann reflectivity~\cite{Oshita:2019sat,Wang:2019rcf}.

Black holes are NH systems. Such NH systems exist in many fields of physics~\cite{Ashida:2020dkc}. A hallmark of NH systems is the existence of exceptional point (EP)~\cite{Heiss:1999alz,Heiss:2012dx,Ryu:2025gnn,Zhong:2018hmw,Dietz:2010bvm}, defined as a spectral singularity where both eigenvalues and their corresponding eigenvectors coalesce. For more mathematical aspects of EPs, we can refer to monograph~\cite{kato1976perturbation}. For the gravity aspects, EPs also appear in the QNM framework of black hole perturbation systems. For massive scalar perturbation of a Kerr black hole, two QNM spectra can become degenerate~\cite{Cavalcante:2024swt,Cavalcante:2024kmy,Cavalcante:2025abr}, where the EP is $(M\mu)_c\simeq0.3704981$ and $(a/M)_c\simeq0.9994660$. Such EP is also obtained in~\cite{Chen:2025sbz}. The EPs can also be found when a Gaussian bump is added to the Regge-Wheeler effective potential~\cite{Yang:2025dbn}, where the EP is $\epsilon\simeq0.005$, and $d\simeq15.698$. Near such EPs, the conventional decomposition of the ringdown waveform into a superposition of QNMs may become invalid~\cite{Yang:2025dbn,PanossoMacedo:2025xnf}. A phenomenon related to the concept of exceptional points is called mode repulsion or avoided crossing, which have been widely studied in~\cite{Dias:2021yju,Motohashi:2024fwt,Oshita:2025ibu,Lo:2025njp,Takahashi:2025uwo}. When the dimension of parameters increases, EPs will form exceptional lines (ELs) which widely arises in non-Hermitian systems~\cite{Shen:2018cjc,Yang:2018dzr,Wu:2024wlf}. For the gravity aspect, in~\cite{Cao:2025afs}, EL is found, and the topological structures (such Berry phase and vorticity) of QNMs associated with EL are also considered. Although we know that the spectrum is unstable, the spectra near exceptional points will be even more unstable, which is studied through the pseudospectra at exceptional points in~\cite{Cao:2025afs}.

In this study, we change the boundary conditions to find exceptional points in the SdS black hole, namely by introducing a complex reflectivity $\mathcal{K}$. In principle, EPs could also be searched for in other two-parameter spaces, e.g. by varying $(\mathcal{K}\in\mathbb{R},\, r_c)$, where $r_c$ corresponds to the cosmological constant (refer to Eqs. (\ref{mass_and_cosmological_constant})). However, we do not adopt it because changing $r_c$ modifies the background spacetime (i.e. the cosmological constant), whereas our goal is to isolate EPs as an intrinsic effect induced by the semi-open boundary condition itself. As far as we know, it is the first work of EPs associated with boundary condition changing in the framework of gravity aspect. More specifically, the reflectivity $\mathcal{K}$ can be prompted into a complex parameter, $\mathcal{K}\in\mathbb{C}$, thus provides two variable real parameters. The existence of imaginary parts in reflectivity means that when waves are reflected at the boundary around the event horizon, not only does their amplitude change, but their phase varies as well. This phase change thereby systematically shifting its QNM spectra.

The work is organized as follows. In Sec. \ref{set_up}, we solve the perturbation equation of SdS black hole by the Heun function, and give the condition of the QNM spectra as the reflectivity exists. In Sec. \ref{QNMs}, the migration of the QNM spectra with respect to the reflectivity in the near-extremal and near-Schwarzschild cases is studied, in which the reflectivity $\mathcal{K}$ restricts to be real and $\mathcal{K}\in[0,1]$. In Sec. \ref{graybody_factors}, the influence of reflectivity on the greybody factor is explored. In Sec. \ref{exceptional_points}, considering that reflectivity is a complex constant, $\mathcal{K}\in\mathbb{C}$, we find out the exception point. Sec. \ref{conclusions} is the conclusions and discussion. In Appendix \ref{Heun_functions}, the process of using the Heun functions to solve QNMs in the SdS black hole is shown. Four expressions of amplitude functions, $A_{\text{in}}(\omega)$, $A_{\text{out}}(\omega)$, $B_{\text{in}}(\omega)$ and $B_{\text{out}}(\omega)$, are given in Appendix \ref{expressions_Ain_Aout_Bin_Bout}. A concrete result of QNMs for $s=0$ is shown in Appendix \ref{QNMs_s_0}.

\section{Sets up for Schwarzschild-de Sitter black holes}\label{set_up}
We start with the line element for Schwarzschild-de Sitter (SdS) black hole. The SdS black hole is described by the metric
\begin{eqnarray}\label{metric_and_function_f}
    \mathrm{d}s^2=-f(r)\mathrm{d}t^2+\frac{\mathrm{d}r^2}{f(r)}+r^2(\mathrm{d}\theta^2+\sin^2\theta\mathrm{d}\phi^2)\, ,\quad f(r)=1-\frac{2M}{r}-\frac{\Lambda r^2}{3}=-\Lambda\frac{(r-r_c)(r-r_e)(r-r_n)}{3r}\, ,
\end{eqnarray}
where the function $f(r)$ has been parameterized by the event horizon $r_e$ and the cosmological horizon $r_c$, and $r_n$ is the third real root of $f(r)$ with $r_n=-(r_e+r_c)<0$. From Eq. (\ref{metric_and_function_f}), one can directly derive the relation between the cosmological constant $\Lambda>0$ and the black hole mass $M$ and horizons via~\cite{Konoplya:2022xid,Zhou:2025xta}
\begin{eqnarray}\label{mass_and_cosmological_constant}
    M=\frac{r_cr_e(r_c+r_e)}{2(r_c^2+r_cr_e+r_e^2)}\, ,\quad \text{and}\quad
    \Lambda=\frac{3}{r_e^2+r_er_c+r_c^2}\, .
\end{eqnarray}
The perturbation equation on this black hole can be reduced to the Schr\"{o}dinger wave-like equations
\begin{eqnarray}\label{master_equation}
    \Big[\frac{\mathrm{d}^2}{\mathrm{d}x^2}+\omega^2-V_s(x
    )\Big]\Psi(x)=0\, ,\quad \mathrm{d}x=\frac{\mathrm{d}r}{f(r)}\, ,
\end{eqnarray}
with the tortoise coordinate $x$ being 
\begin{eqnarray}\label{tortoise_coordinates_integration}
    x=\frac{3r_e}{\Lambda(r_c-r_e)(r_e-r_n)}\ln\Big|1-\frac{r_e}{r}\Big|+\frac{3r_c}{\Lambda(r_e-r_c)(r_c-r_n)}\ln\Big|1-\frac{r_c}{r}\Big|+\frac{3r_n}{\Lambda(r_e-r_n)(r_n-r_c)}\ln\Big|1-\frac{r_n}{r}\Big|\, .
\end{eqnarray}
Here, we consider three kinds of potentials whose expressions are~\cite{Zhidenko:2003wq,Arnaudo:2025kit}
\begin{eqnarray}\label{potentials}
    V_{s}(r)=f(r)\Big[\frac{\ell(\ell+1)}{r^2}+(1-s)\frac{f^{\prime}(r)}{r}+\delta_{s,2}f^{\prime\prime}(r)+\frac{2\Lambda}{3}\delta_{s,0}\Big]\, ,\quad \text{for} \quad s=0\, ,1\, ,2\, ,
\end{eqnarray}
where $s=0,1,2$ indicate scalar, electromagnetic and gravitational axial perturbations, respectively, and $\delta_{s,2}$, $\delta_{s,0}$ stand for the Kronecker delta. Regarding the scalar perturbation $s = 0$, it should be noted that the case discussed here is the conformally coupled massless scalar perturbation~\cite{Crispino:2013pya}, $(\square-R/6)\Phi=0$, where for the Schwarzschild-de Sitter black hole the scalar curvature is satisfied with $R=4\Lambda$. In the case of other types of perturbations, such as the minimally coupled massless scalar, the corresponding differential equation does not simplify to a Heun’s equation. Mathematically, for the minimally coupled massless scalar, although the infinity point $r=\infty$ is a regular singular point, the difference between the two associated indicial exponents is $3$. One can use the so-called Liouville invariant to demonstrate that, with this specific structure, this regular singular point can not be removed. 

Like the P\"{o}schl-Teller potential~\cite{Solidoro:2024yxi}, the analytical solution for the potential (\ref{potentials}) of the Schwarzschild-de Sitter black hole can be derived. From Appendix \ref{Heun_functions}, at the event horizon $z=0$, one finds the general local solution $\Psi(r)$ which can be written as
\begin{eqnarray}\label{general_solution_z_0}
    \Psi(r)=A\cdot \Psi_{01}(z)+B\cdot \Psi_{02}(z)\, ,\quad z=\frac{r_c(r-r_e)}{(r_c-r_e)r}\, ,
\end{eqnarray}
where $A$, $B$ are general constants, which are just used to indicate a general solution. The function $\Psi_{01}$ is an outgoing solution at the event horizon, namely
\begin{eqnarray}\label{Psi_01}
    \Psi_{01}(z)=z^{\rho_{e,1}}(z-1)^{\rho_{c,1}}(z-z_n)^{\rho_{n,1}}Hl(z_n,q;\alpha,\beta,\gamma,\delta;z)\, ,
\end{eqnarray}
while the function $\Psi_{02}$ is an ingoing solution at the event horizon, namely
\begin{eqnarray}\label{Psi_02}
    \Psi_{02}(z)=z^{\rho_{e,1}}(z-1)^{\rho_{c,1}}(z-z_n)^{\rho_{n,1}}z^{1-\gamma}Hl(z_n,(z_n\delta+\epsilon)(1-\gamma)+q;\alpha+1-\gamma,\beta+1-\gamma,2-\gamma,\delta;z)\, .
\end{eqnarray}
The symbol $Hl$ standing for ``Heun-local'' corresponds to the Heun function, which is a built-in function written ``HeunG'' in \textit{Mathematica}.  At the cosmological horizon $z=1$, one finds the general local solution $\Psi(r)$ can be written as
\begin{eqnarray}\label{general_solution_z_1}
    \Psi(r)=A\cdot \Psi_{11}(z)+B\cdot \Psi_{12}(z)\, ,\quad z=\frac{r_c(r-r_e)}{(r_c-r_e)r}\, ,
\end{eqnarray}
where the function $\Psi_{11}$ is an outgoing solution at the cosmological horizon, namely
\begin{eqnarray}\label{Psi_11}
    \Psi_{11}(z)=z^{\rho_{e,1}}(z-1)^{\rho_{c,1}}(z-z_n)^{\rho_{n,1}}Hl(1-z_n,\alpha\beta-q;\alpha,\beta,\delta,\gamma;1-z)\, ,
\end{eqnarray}
while the function $\Psi_{12}$ is an ingoing solution at the cosmological horizon, namely
\begin{eqnarray}\label{Psi_02}
    \Psi_{12}(z)&=&z^{\rho_{e,1}}(z-1)^{\rho_{c,1}}(z-z_n)^{\rho_{n,1}}(1-z)^{1-\delta}\nonumber\\
    &&\times Hl(1-z_n,((1-z_n)\gamma+\epsilon)(1-\delta)+\alpha\beta-q;\alpha+1-\delta,\beta+1-\delta,2-\delta,\gamma;1-z)\, .
\end{eqnarray}
Note that up to now, we have not imposed any boundary conditions.

When solving two-point boundary value problems between $z=0$ and $z=1$, a key step involves determining how the local solutions in various domains connect to one another. The relation between $\{u_{01},u_{02}\}$ and $\{u_{11},u_{12}\}$ is
\begin{eqnarray}\label{relations_u}
    u_{11}(z)&=&C_{11}u_{01}(z)+C_{12}u_{02}(z)\, ,\nonumber\\
    u_{12}(z)&=&C_{21}u_{01}(z)+C_{22}u_{02}(z)\, ,
\end{eqnarray}
where the connection coefficients read
\begin{eqnarray}\label{functions_C}
    C_{11}=\frac{W[u_{11},u_{02}]}{W[u_{01},u_{02}]}\, ,\quad C_{12}=\frac{W[u_{11},u_{01}]}{W[u_{02},u_{01}]}\, ,\quad C_{21}=\frac{W[u_{12},u_{02}]}{W[u_{01},u_{02}]}\, ,\quad C_{22}=\frac{W[u_{12},u_{01}]}{W[u_{02},u_{01}]}\, ,
\end{eqnarray}
and the Wronskian is defined as $W[u,v]\equiv u\mathrm{d}v/\mathrm{d}z-v\mathrm{d}u/\mathrm{d}z$. Conversely, the inverse relation is
\begin{eqnarray}\label{relations_u_inverse}
    u_{01}(z)&=&D_{11}u_{11}(z)+D_{12}u_{12}(z)\, ,\nonumber\\
    u_{02}(z)&=&D_{21}u_{11}(z)+D_{22}u_{12}(z)\, ,
\end{eqnarray}
where the connection coefficients read
\begin{eqnarray}\label{functions_D}
    D_{11}=\frac{W[u_{01},u_{12}]}{W[u_{11},u_{12}]}\, ,\quad D_{12}=\frac{W[u_{01},u_{11}]}{W[u_{12},u_{11}]}\, ,\quad D_{21}=\frac{W[u_{02},u_{12}]}{W[u_{11},u_{12}]}\, ,\quad D_{22}=\frac{W[u_{02},u_{11}]}{W[u_{12},u_{11}]}\, .
\end{eqnarray}
The four functions $u_{01}(z)$, $u_{02}(z)$, $u_{11}(z)$ and $u_{12}(z)$ are from Eqs. (\ref{u01_u02}) and Eqs. (\ref{u11_u12}) in Appendix \ref{Heun_functions}. It is not difficult to prove that the ratio of the Wronskian is independent of $z$ from which the connection coefficients in Eqs. (\ref{functions_C}) and Eqs. (\ref{functions_D}) are independent of $z$. Therefore, in the evalution of the above Wronskians, one can choose a value of $z$ located in both convergence circles at $z=0$ and $z=1$. In practice, the middle ponit $z_0=1/2$ is good point to achieve this purpose.

The standard ``in'' and ``up'' solutions of Eq. (\ref{master_equation}) are defined as
\begin{eqnarray}\label{in_solution}
    \Psi_{\text{in}}(x)=
    \left\{
    \begin{array}{l}
    \mathrm{e}^{-\mathrm{i}\omega x}\, ,\quad x\to-\infty\\
    A_{\text{in}}(\omega)\mathrm{e}^{-\mathrm{i}\omega x}+A_{\text{out}}(\omega)\mathrm{e}^{\mathrm{i}\omega x}\, ,\quad x\to+\infty
    \end{array}\right.\, ,
\end{eqnarray}
and 
\begin{eqnarray}\label{up_solution}
    \Psi_{\text{up}}(x)=
    \left\{
    \begin{array}{l}
    B_{\text{in}}(\omega)\mathrm{e}^{-\mathrm{i}\omega x}+B_{\text{out}}(\omega)e^{\mathrm{i}\omega x}\, ,\quad x\to-\infty\\
    \mathrm{e}^{\mathrm{i}\omega x}\, ,\quad x\to+\infty
    \end{array}\right.\, .
\end{eqnarray}
The detailed expressions of $A_{\text{in}}(\omega)$, $A_{\text{out}}(\omega)$, $B_{\text{in}}(\omega)$ and $B_{\text{out}}(\omega)$ are found in Appendix \ref{expressions_Ain_Aout_Bin_Bout}. Consider the semi-open problem in which a wall of reflectivity $\mathcal{K}(\omega)$ is located at $x_0\ll0$ such that $V_s(x_0)$ is approximated as zero, and the solution $\Psi_s(x)$ at $x=x_0$ is a linear combination of ingoing and outgoing waves $\mathrm{e}^{\pm\mathrm{i}\omega x}$. To give a clear description of the proximity between the wall and the event horizon, we convert the wall position $x_0$ into the corresponding areal radial coordinate $r_0$ and use $\Delta\equiv(r_0-r_e)/r_e$ to portray. For such a situation, the solution of Eq. (\ref{master_equation}) obeys the following boundary condition~\cite{Berti:2025hly,Macedo:2018yoi,Solidoro:2024yxi,Rosato:2025byu,Herdeiro:2019fps}
\begin{eqnarray}\label{semi_open_conditions}
    \Psi_s(x)= \left\{
    \begin{array}{l}
    \mathrm{e}^{-\mathrm{i}\omega x}+\mathcal{K}(\omega)\mathrm{e}^{-2\mathrm{i}\omega x_0} \mathrm{e}^{\mathrm{i}\omega x}\, ,\quad x\to x_0\\
    S_{\text{in}}(\omega)\mathrm{e}^{-\mathrm{i}\omega x}+S_{\text{out}}(\omega)\mathrm{e}^{\mathrm{i}\omega x}\, ,\quad x\to+\infty
    \end{array}\right.\, .
\end{eqnarray}
From the property of being constant for the Wronskians of two solutions of Eq. (\ref{master_equation}), one can write the following three useful relations between the coefficients $A_{\text{in}}(\omega)$, $A_{\text{out}}(\omega)$, $B_{\text{in}}(\omega)$, $B_{\text{out}}(\omega)$, $S_{\text{in}}(\omega)$, and $S_{\text{out}}(\omega)$:
\begin{eqnarray}\label{Relation}
    A_{\text{in}}(\omega)=B_{\text{out}}(\omega)\, ,\quad S_{\text{out}}(\omega) A_{\text{in}}(\omega)-S_{\text{in}}(\omega) A_{\text{out}}(\omega)=\mathcal{K}(\omega)\mathrm{e}^{-2\mathrm{i}\omega x_0}\, ,\quad S_{\text{in}}(\omega)=B_{\text{out}}(\omega)-B_{\text{in}}(\omega)\mathcal{K}(\omega)\mathrm{e}^{-2\mathrm{i}\omega x_0}\, ,
\end{eqnarray}
where $\omega\in\mathbb{C}$. Therefore, for the semi-open problem, the QNM spectra are defined by the resonance condition $S_{\text{in}}(\omega)=0$. For this time, the solution $\Psi_s(x)$ is outgoing at the cosmological horizon satisfying the boundary condition. Note that for the open system, i.e., $\mathcal{K}(\omega)=0$, the condition of the QNMs reduces to $B_{\text{out}}(\omega)=C_{11}(\omega)=0$. 

\section{Quasinormal modes}\label{QNMs}
In this section, we will study the QNM spectra for the existence of a wall of reflectivity $\mathcal{K}(\omega)$ in the Schwarzschild-de Sitter black hole. Here, we always focus on the axial gravitational perturbation, i.e., $s=2$, and the angular momentum parameter $\ell$ is set to be $2$. Perturbations of other spins satisfy the same master equation with similar effective potentials, consequently, the qualitative conclusions drawn here should remain valid for them (One can find a special case of QNMs for $s=0$ in Appendix \ref{QNMs_s_0}.). For the sake of simplicity, although $\mathcal{K}(\omega)$ may depend on $\omega$, in this section, we focus on the reflectivity that does not depend on the frequency. At this time, one can write $\mathcal{K}(\omega)$ as $\mathcal{K}\in\mathbb{C}$ with $|\mathcal{K}|\le1$. This condition allow us to vary the amount of the absorption for the wall. To be more specific, for $\mathcal{K}=1$, the wall is fully reflective, namely the Neumann boundary condition, while for $\mathcal{K}=0$, it recovers open boundary condition. Note that for $\mathcal{K}=-1$, we will get the Dirichlet boundary condition. More importantly, as $\mathcal{K}$ has an imaginary part (Note that this case will be studied associated with the EPs in Sec. \ref{exceptional_points}), such wall is going to lead the Robin boundary condition~\cite{Solidoro:2024yxi,Herdeiro:2019fps}.

We are concerned with the most common scenario where $\mathcal{K}\in\mathbb{R}$ satisfies $0\le\mathcal{K}\le1$. The QNM spectra of the semi-open system for the SdS black hole are obtained by numerically solving $S_{\text{in}}(\omega)=0$, namely QNM spectra are the zeros of the function $S_{\text{in}}(\omega)$. For a non-zero reflectivity, one should expect the QNM spectra will have a deviation from those of the original open problem. The problem of finding the zeros of nonlinear functions usually requires reliable iterative initial values, and Ref. \cite{Wu:2025sbq} provides a practical and feasible way to find them. However, in order to capture the essence of the impact of reflectivity on QNM spectra, we consider two scenarios, namely the near extreme SdS case and the near Schwarzschild case. For these two cases, one can use the QNM spectra results from near extreme SdS case and the Schwarzschild case as the iterative initial values for getting the zeros of the function $S_{\text{in}}(\omega)$ within $\mathcal{K}=0$. For the near extremal SdS black hole, which is defined as the spacetime for which the cosmological horizon $r_c$ is very close to the black hole horizon $r_e$, i.e., $r_e/r_c\approx1$, the analytical expressions of the QNM spectra have been found in~\cite{Cardoso:2003sw,Konoplya:2004uk,Jansen:2017oag}. For $s=2$, the expression is given by
\begin{eqnarray}
    \omega_n=\kappa_e\Bigg[-\Big(n+\frac12\Big)\mathrm{i}+\sqrt{(\ell+2)(\ell-1)-\frac14}\Bigg]\, ,\quad n=0,1,\cdots\, ,
\end{eqnarray}
where $\kappa_e$ is the surface gravity at the event horizon. The simple form of the QNM spectra in extreme SdS black holes is attributed to the behavior of the effective potential approximating the P\"{o}shl-Teller potential~\cite{Ferrari:1984zz}. For the Schwarzschild black hole, one can find QNM spectra in~\cite{Berti:2009kk,Mamani:2022akq,Leaver:1985ax}.

\begin{figure}[htbp]
    \centering
   \subfigure[]{\includegraphics[width=0.45\linewidth]{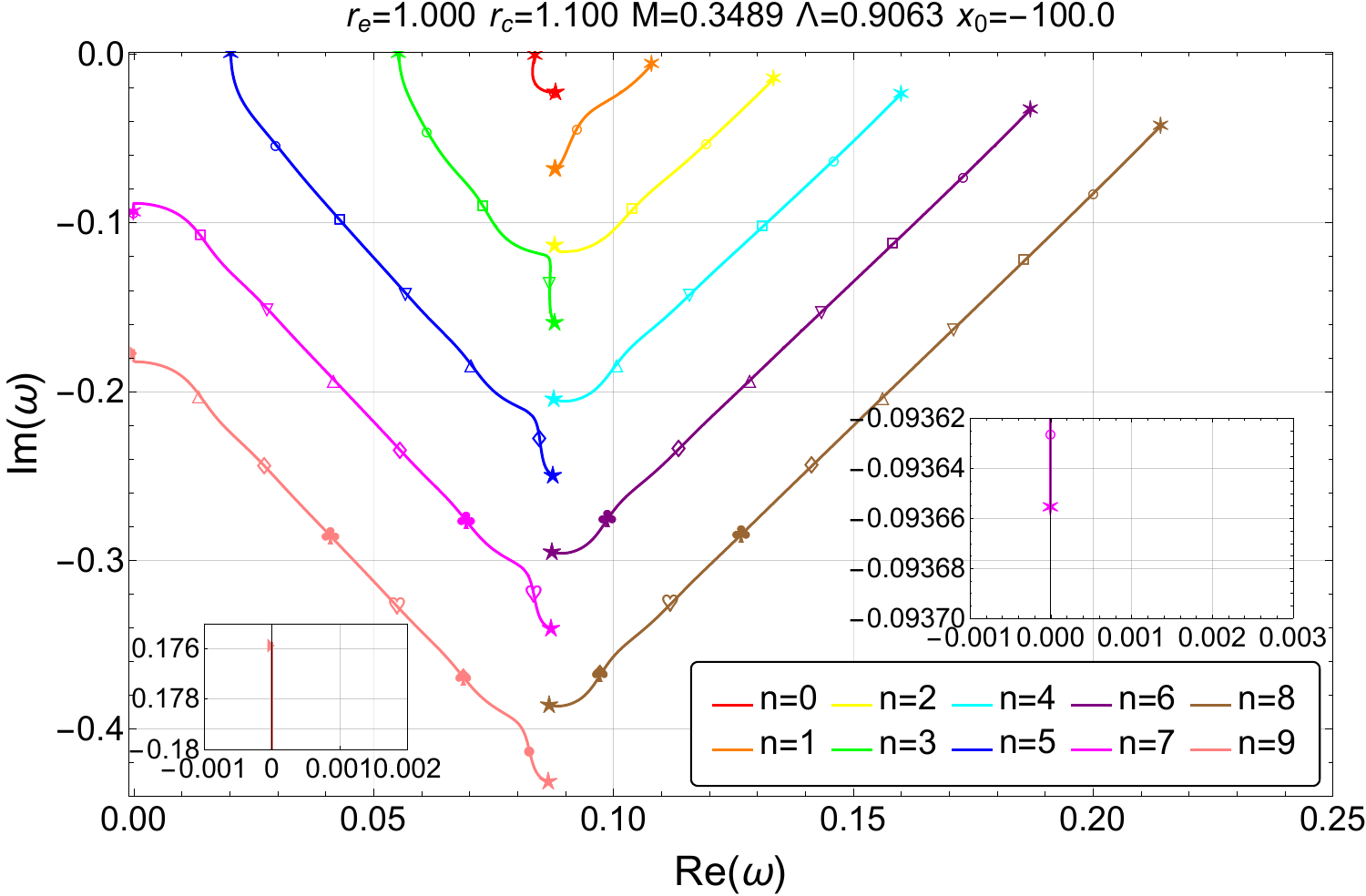}\label{QNM_Spectra_near_extreme_x0_m_100}}\hfill
   \subfigure[]{\includegraphics[width=0.45\linewidth]{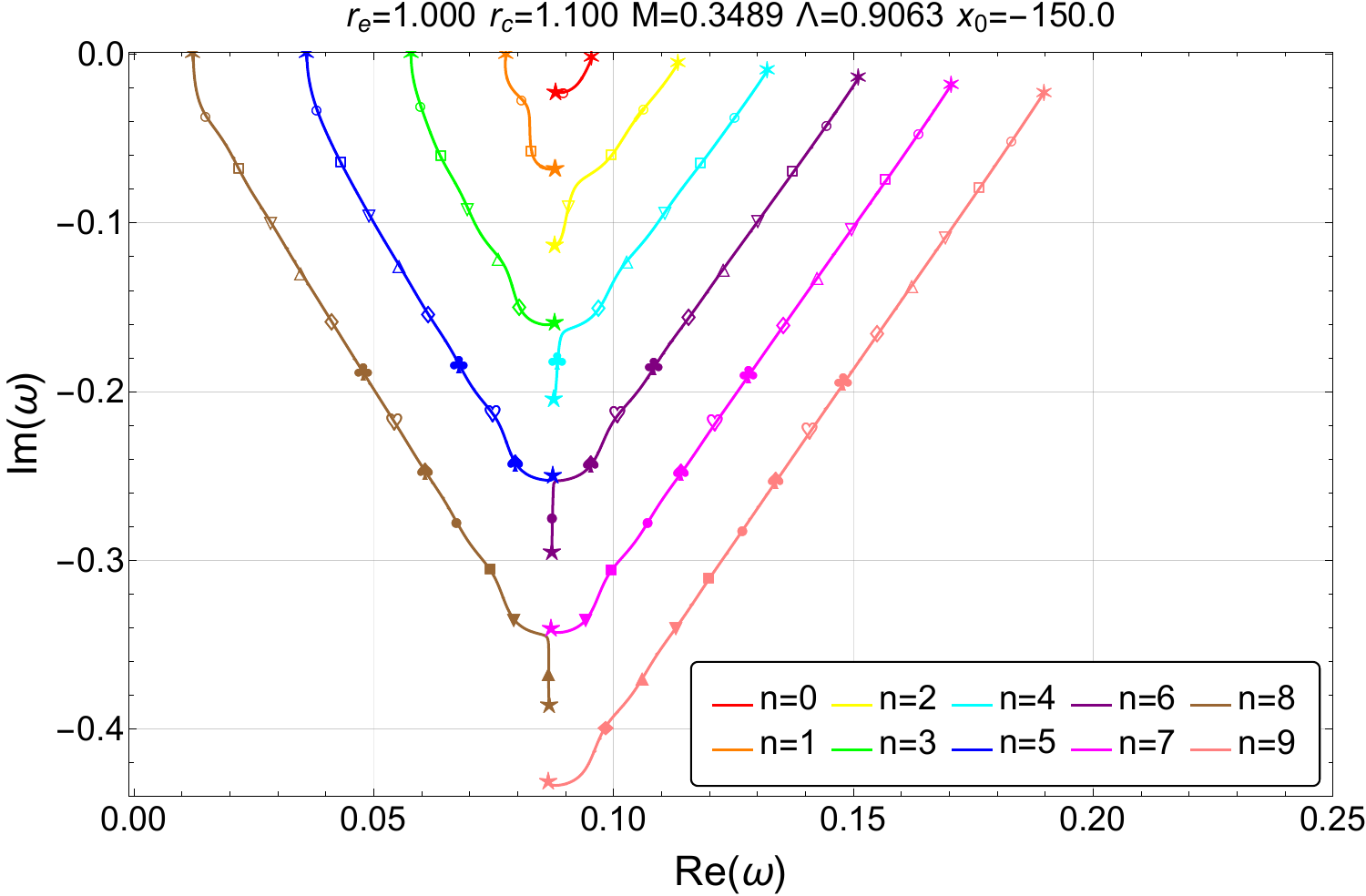}\label{QNM_Spectra_near_extreme_x0_m_150}}
    \caption{The migrations of the QNM spectra with $\mathcal{K}\in[0,1]$ varying in the complex plane for the first $10$ modes, where the parameters of panel (a) are chosen $r_e=1$, $r_c=1.1$, and $x_0=-100$ $(\Delta=6.8341915\times10^{-6})$ while the parameters of panel (b) are chosen $r_e=1$, $r_c=1.1$, and $x_0=-150$ $(\Delta=6.3243116\times10^{-8})$. Different color lines correspond to different modes. Dots of different shapes represent $\mathcal{K}$ of different intensities, where $\star=0$, $\otimes=10^{-60}$, $\oplus=10^{-56}$, $\blacklozenge=10^{-52}$, $\blacktriangle=10^{-48}$, $\blacktriangledown=10^{-44}$, $\blacksquare=10^{-40}$, $\bullet=10^{-36}$, $\spadesuit=10^{-32}$, $\heartsuit=10^{-28}$, $\clubsuit=10^{-24}$, $\diamondsuit=10^{-20}$, $\triangle=10^{-16}$, $\triangledown=10^{-12}$, $\square=10^{-8}$, $\circ=10^{-4}$, and $\ast=1$, the same below. Two subfigures in left panel are the zoom-in near the imaginary axis for the modes $n=7$ and $n=9$. The symbol $\blacktriangleright$ with pink color in the left panel stands for the parameter $\mathcal{K}=3.57\times 10^{-14}$.}
    \label{QNM_Spectra_near_extreme}
\end{figure}

In Fig. \ref{QNM_Spectra_near_extreme} (near extreme case), we display their continuous migrations in the complex plane as $\mathcal{K}$ is varied, fixing $r_e=1$ and $r_c=1.1$, where in Fig. \ref{QNM_Spectra_near_extreme_x0_m_100}, the reflectivity wall is located at $x_0=-100$ while in Fig. \ref{QNM_Spectra_near_extreme_x0_m_150} the reflectivity wall is located at $x_0=-150$. A total of $10$ mode migrations are presented. The mark $\star$ represents the QNM spectra of the open system, i.e., $\mathcal{K}=0$. As $\mathcal{K}$ is increased, the modes migrate down toward the real line. Intuitively, it can be identified that three qualitatively different types of QNMs, which is also emphasized and explained in recent work~\cite{Solidoro:2024yxi}. First type of modes migrate to the real axis (in fact, the imaginary parts of them are very small), for example $n=0$, $n=3$, $n=5$ in Fig. \ref{QNM_Spectra_near_extreme_x0_m_100} with $n=1$, $n=3$, $n=5$, $n=8$ in Fig. \ref{QNM_Spectra_near_extreme_x0_m_150}. These modes become the quasibound states (QBS) which have long lifetimes. This behavior is naturally understood as follow. The real part $\text{Re}(\omega_0)$ serves as a practical threshold for traversing the potential barrier, where $\omega_0$ is the original fundamental QNM. These modes with $0<\text{Re}(\omega)<\text{Re}(\omega_0)$ lie effectively below this threshold, cannot propagate across the photon-sphere region, and are therefore trapped within the wall--barrier cavity, escaping only via tunneling through the effective potential barrier, which naturally yields very small imaginary parts and long lifetimes. The second branch of modes, although also gradually approaching the real axis, do not have an imaginary part that is entirely zero when $\mathcal{K}=1$ for example $n=1$, $n=2$, $n=4$, $n=6$, $n=8$ in Fig. \ref{QNM_Spectra_near_extreme_x0_m_100} with $n=0$, $n=2$, $n=4$, $n=6$, $n=7$, $n=9$ in Fig. \ref{QNM_Spectra_near_extreme_x0_m_150}. Which correspond to the over-barrier regime, $\text{Re}(\omega)>\text{Re}(\omega_0)$: they exceed the effective energy threshold for traversing the photon-sphere potential barrier, therefore they effectively pass over the barrier, experience only weak reflection of potential and escape to infinity. Consequently, they are more strongly damped, retaining a finite decay rate and can be identified with the ``new overtones'' of the semi-open system. Last, the third type of mode will migrate to the imaginary axis as $\mathcal{K}$ increases, becoming $\text{Re}(\omega)=0$, for example $n=7$, $n=9$ in Fig. \ref{QNM_Spectra_near_extreme_x0_m_100}. These modes are caused by the presence of attractors from the perspective of dynamical system~\cite{Torres:2026uey}. Although they study a localized modification of the effective potential of the form $\epsilon\delta(x-x_0)$, rather than the changing the boundary condition considered here, these two approaches are physically equivalent in the relevant limit: as $\epsilon$ increases, the $\epsilon\delta(x-x_0)$ modification acts increasingly like an effective reflective wall. They identify attractors on the imaginary axis, which naturally accounts for the emergence of purely damped modes. Furthermore, comparing the two panels in Fig. \ref{QNM_Spectra_near_extreme}, we can also conclude that the closer $x_0$ is to the event horizon, the closer the modes $(n\ge10)$ are to the real axis when $\mathcal{K}=1$. It can be also concluded that for Fig. \ref{QNM_Spectra_near_extreme_x0_m_150}, those modes that collide with the imaginary axis occur at higher order overtones, although we have not shown. Additionally, as $x_0$ decreases, the instability of the mode spectrum increases, meaning that a smaller $\mathcal{K}$ can cause a significant shift in the spectrum. 

\begin{figure}[htbp]
    \centering
     \subfigure[]{\includegraphics[width=0.45\linewidth]{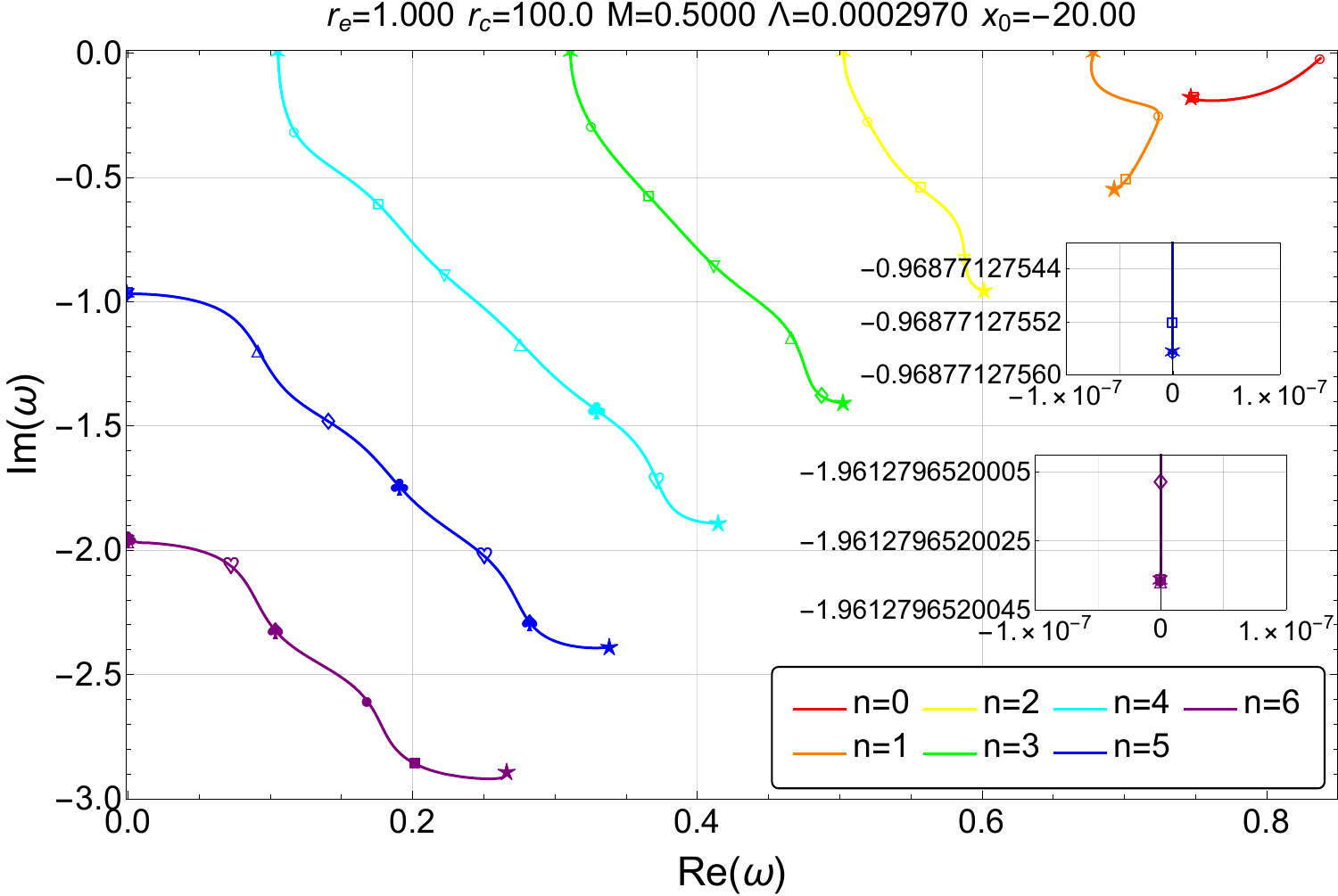}\label{QNM_Spectra_near_Sch_x0_m_20}}\hfill
     \subfigure[]{\includegraphics[width=0.45\linewidth]{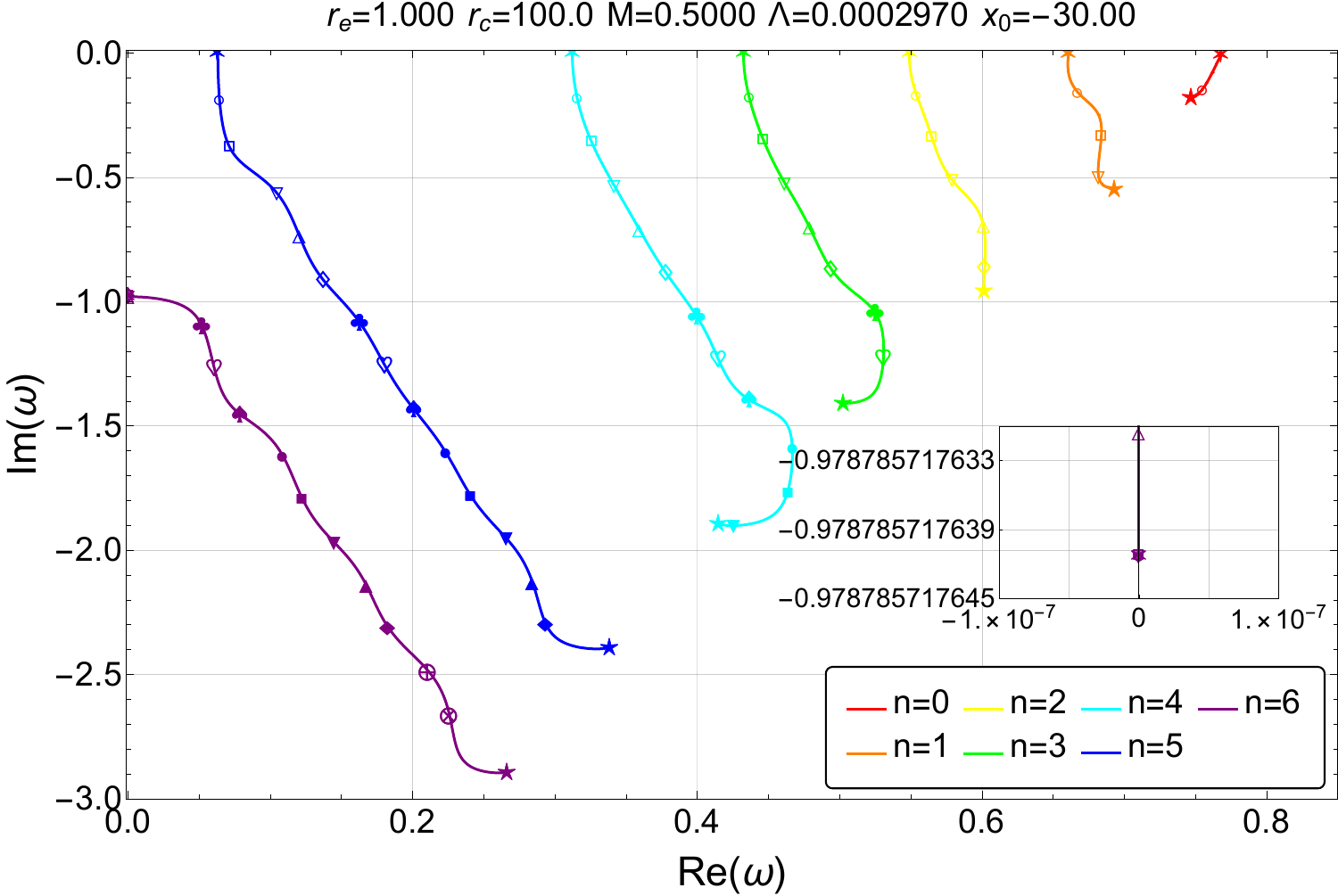}\label{QNM_Spectra_near_Sch_x0_m_30}}
    \caption{The migrations of the QNM spectra with $\mathcal{K}\in[0,1]$ varying in the complex plane for the first $7$ modes, where the parameters of panel (a) are chosen $r_e=1$, $r_c=100$, and $x_0=-20$ $(\Delta=4.6490692\times10^{-8})$ while the parameters of panel (b) are chosen $r_e=1$, $r_c=100$, and $x_0=-30$ $(\Delta=2.1169523\times10^{-12})$. Three subfigures above are the zoom-in near the imaginary axis.}
    \label{QNM_Spectra_near_Sch}
\end{figure}

In Fig. \ref{QNM_Spectra_near_Sch} (near Schwarzschild case), we display their continuous migrations in the complex plane as $\mathcal{K}$ is varied, fixing $r_e=1$ and $r_c=100$, where in Fig. \ref{QNM_Spectra_near_Sch_x0_m_20}, the reflectivity wall is located at $x_0=-20$ while in Fig. \ref{QNM_Spectra_near_Sch_x0_m_30} the reflectivity wall is located at $x_0=-30$. In terms of qualitative properties, the migration situation near Schwarzschild case is similar to the migration situation near extremes case. As $\mathcal{K}$ increases, the QNM spectra approach the real and imaginary axes. Therefore, there will be no further elaboration.

\section{Greybody factors}\label{graybody_factors}
Greybody factors encode the frequency-dependent transmission of perturbations across the effective potential barrier of a black hole spacetime. They enter directly in the computation of Hawking fluxes and energy spectra. Unlike individual QNM spectra, they often behave as remarkably robust observables~\cite{Oshita:2023cjz,Rosato:2024arw,Oshita:2024fzf,Ianniccari:2024ysv,Wu:2024ldo,Kyutoku:2022gbr,Konoplya:2025ixm,Wu:2025sbq,Xie:2025jbr}. The robustness of the greybody factors indicates that ringdown waveform built directly from the greybody factors may offer a better modelling strategy in future gravitational wave studies. Motivated by possible quantum gravity effects, a variety of proposals for horizon scale structure (ranging from firewalls~\cite{Bousso:2025udh} and fuzzballs~\cite{Mathur:2008kg} to more general exotic compact objects~\cite{Mark:2017dnq}) suggest that the classical picture of a perfectly absorbing event horizon may break down at the would-be horizon. We therefore consider a semi-open system, in which the near horizon region is modelled by a partially reflecting surface.

In this section, we focus on the greybody factors (transmission coefficients) of the Schwarzschild-de Sitter black hole with a semi-open condition to account for some possible quantum corrections near the event horizon. We consider a setup in which a partially reflecting surface is introduced just outside the black hole horizon. The wall is placed at a fixed tortoise coordinate $x_0 \ll 0$, in a region where the effective potential can be approximated as $V_s(x_0) \simeq 0$. Two representative reflection models are considered: a frequency-independent (constant) reflectivity~\cite{Mark:2017dnq} and a Boltzmann-type reflectivity~\cite{Wang:2019rcf,Oshita:2019sat}. Under the boundary condition (\ref{semi_open_conditions}) and the definition of greybody factor~\cite{Rosato:2025byu},
\begin{eqnarray}
    R(\omega)\equiv \Bigg|\frac{S_{\text{out}}(\omega)}{S_{\text{in}}(\omega)}\Bigg|^2\, ,
    \qquad \Gamma(\omega)\equiv 1-R(\omega)\, ,\qquad \omega\in\mathbb{R}\, ,
\end{eqnarray}
one can obtain the analytic expression for the greybody factor using the relations in Eqs. (\ref{Relation}),
\begin{eqnarray}\label{GFs}
    \Gamma(\omega) =1-\Bigg|\frac{1}{A_{\text{in}}(\omega)}\Big[A_{\text{out}}(\omega)+\frac{\mathcal{K}(\omega)\mathrm{e}^{-2\mathrm{i}\omega x_0}}{B_{\text{out}}(\omega)-B_{\text{in}}(\omega)\mathcal{K}(\omega)\mathrm{e}^{-2\mathrm{i}\omega x_0}}\Big]\Bigg|^2\, .
\end{eqnarray}
In the followings, we will compute $\Gamma(\omega)$ and $R(\omega)$ for different reflectivities $\mathcal{K}(\omega)$ and wall positions $x_0$.

We first consider a frequency-independent reflectivity,
\begin{equation}
    \mathcal{K}(\omega) = C\, ,
    \qquad 0 \leq C \leq 1\, .
    \label{eq:GF_K_const}
\end{equation}
It is the simplest phenomenological model which commonly adopted in the ECO, where such frequency-independent reflectivity has been used to get the QNM spectra in Sec. \ref{QNMs}.

\begin{figure}[htbp]
    \centering
    \subfigure[]
     {\includegraphics[width=0.45\linewidth]{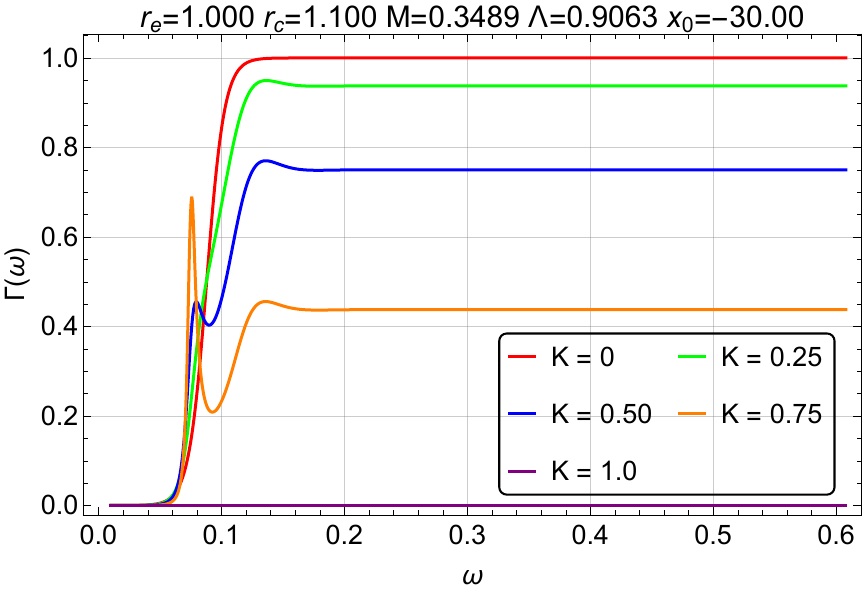}}\hfill
    \subfigure[]
    {\includegraphics[width=0.45\linewidth]{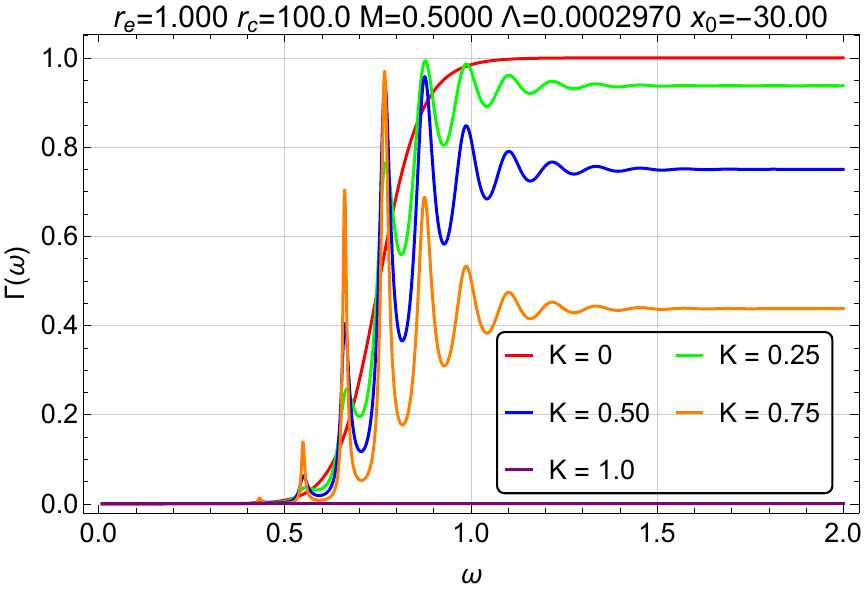}}
    \caption{Greybody factor $\Gamma(\omega)$ for axial gravitational $(s=2)$ perturbations of the Schwarzschild-de Sitter black hole with a frequency-independent reflectivity $\mathcal{K}$. In both panels the reflective wall is placed at $x_0=-30$ ($\Delta=4.5730021\times10^{-3}$ and $\Delta=2.1169523\times10^{-12}$), and the curves with different colors correspond to $\mathcal{K}=0,\,0.25,\,0.50,\,0.75,$ and $1$. Panel (a) shows the near-extremal case with $r_e=1$ and $r_c=1.1$, and panel (b) shows the near Schwarzschild case with $r_e=1$ and $r_c=100$.}
    \label{GFsVaryK}
\end{figure}

\begin{figure}[htbp]
    \centering
    \subfigure[]
     {\includegraphics[width=0.45\linewidth]{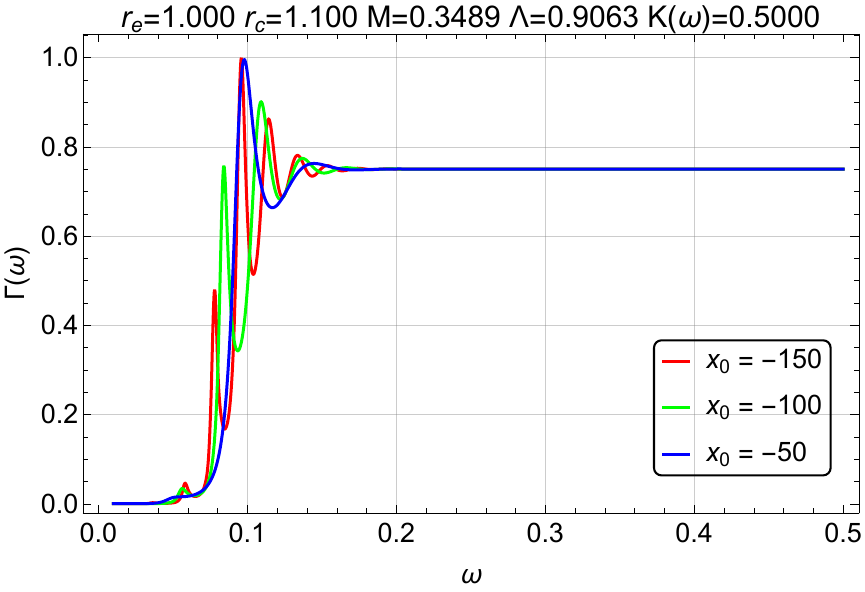}}\label{GFs_Vary_x0_rc_11over10}\hfill
    \subfigure[]
    {\includegraphics[width=0.45\linewidth]{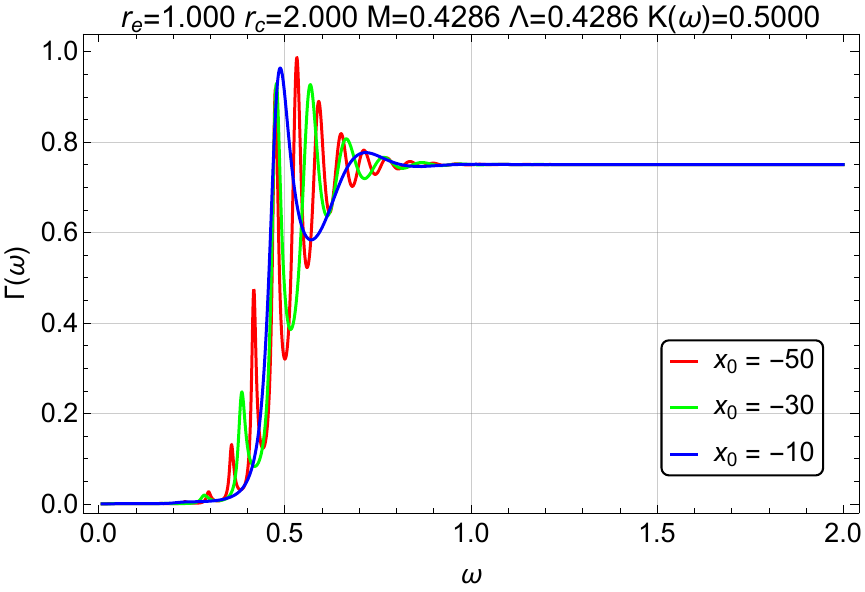}}\label{GFs_Vary_x0_rc_2}
    \subfigure[]
    {\includegraphics[width=0.45\linewidth]{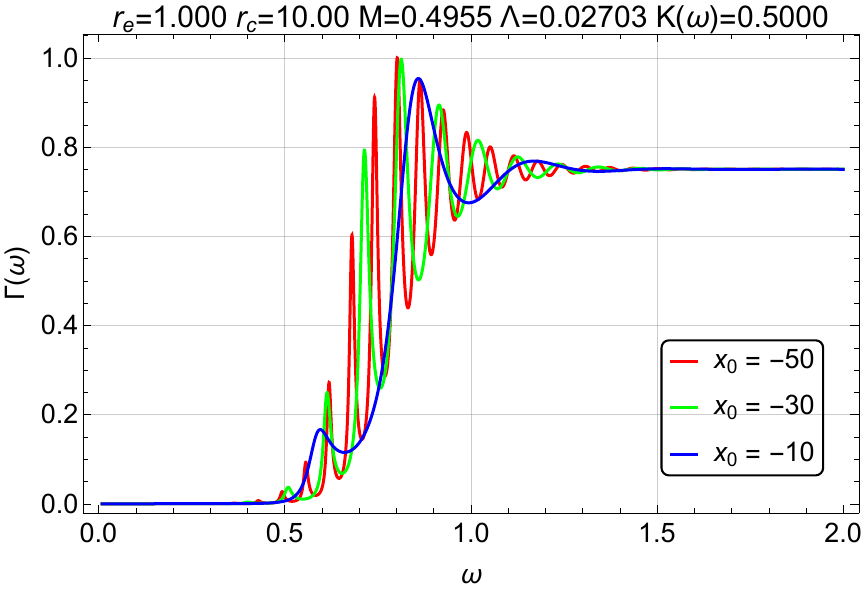}}\label{GFs_Vary_x0_rc_10}\hfill
    \subfigure[]
    {\includegraphics[width=0.45\linewidth]{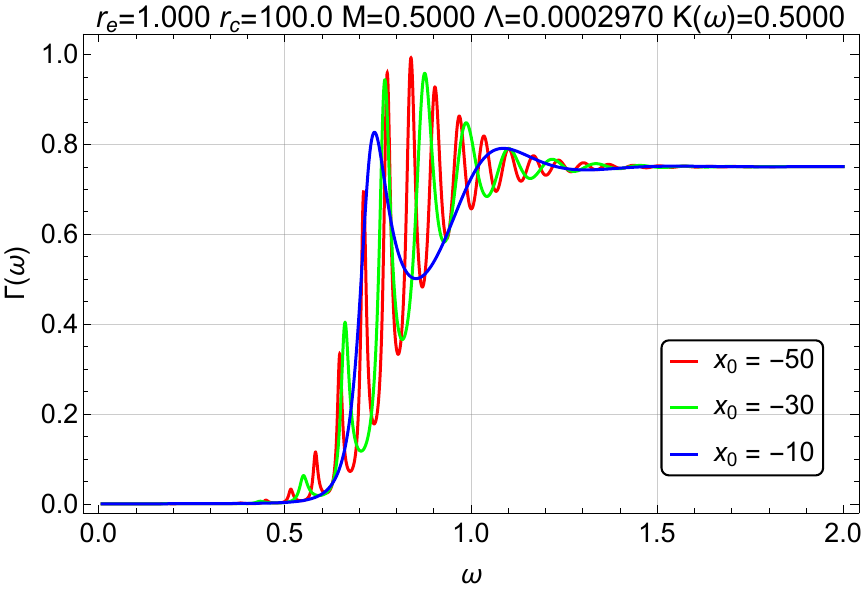}}\label{GFs_Vary_x0_rc_100}
    \caption{Greybody factor $\Gamma(\omega)$ for axial gravitational $(s=2)$ perturbations of the Schwarzschild-de Sitter black hole with constant reflectivity $\mathcal{K}(\omega)=0.5$ and different positions of the reflective walls. The curves in panel (a) correspond to $x_0=-150$, $-100$, and $-50$ in the near-extremal case with $r_e=1$, $r_c=1.1$. Panels (b), (c), and (d) show respectively, $r_c=2$, $r_c=10$, and $r_c=100$, with $x_0=-50$, $-30$, and $-10$ in each panel.}
    \label{GFsVaryX0}
\end{figure}

In Fig. \ref{GFsVaryK}, we show the greybody factors $\Gamma(\omega)$ for different values of $\mathcal{K}$, at fixed wall position $x_0=-30$ in two Schwarzschild-de Sitter spacetimes. For $\mathcal{K}=0$, there is no reflection at the wall and in this case the greybody factor is a smooth, monotonically increasing curve, which is the usual behaviour found for black holes. For $\mathcal{K}=1$, the system describes complete reflection at the wall, and accordingly the corresponding curve is a horizontal line with $\Gamma(\omega)=0$. From the two panels of Fig. \ref{GFsVaryK} we see that, when $\mathcal{K}$ lies between $0$ and $1$, the greybody factors exhibit clear oscillations. However, as $\omega$ tends to infinity, the greybody factor $\Gamma(\omega)\xrightarrow[\omega\to\infty]{} 1-\mathcal{K}^2$, and the behaviour of the greybody factor is entirely dominated by the reflective wall. In Fig. \ref{GFsVaryX0}, we focus on the dependence of the greybody factor on the position of the reflective wall. We find that, for the same SdS spacetime (i.e. for same $r_e$ and $r_c$), moving the reflective wall closer to the event horizon increases the number of resonance peaks and decreases the spacing between neighbouring peaks.

The origin of these resonances is that, when a reflective surface is introduced near the horizon, the echoes reflected back from the wall interfere with the incident wave and thus give rise to the resonant behaviour~\cite{Solidoro:2024yxi}. The number and positions of the resonant peaks are entirely determined by the effective distance $L_{\rm eff}$ between the wall position and the effective potential~\cite{Solidoro:2024yxi,Rosato:2025byu,Rosato:2025lxb}. When the wall position $x_0$ is fixed and $r_c$ is increased, or when $r_c$ is fixed and the reflective wall $x_0$ is moved closer to the horizon, $L_{\rm eff}$ increases in both cases, and the number of resonant peaks correspondingly increases. The right panel of Fig. \ref{GFsVaryK} clearly shows that, when $L_{\rm eff}$ is fixed, the number and positions of the resonant peaks are almost the same for different values of $\mathcal{K}$, and only the peak amplitudes change, being correlated with $\mathcal{K}$.

We now turn to the case in which the reflectivity is frequency-dependent and obeys a Boltzmann-type condition. This choice is motivated by the idea that possible quantum structures near the event horizon may behave as a dissipative layer at the Hawking temperature $T_\text{H}$. In such a picture, we set
\begin{eqnarray}\label{eq:GF_K_Boltz}
    \mathcal{K}(\omega) = \exp\!\left(-\frac{|\omega|}{T_\text{H}}\right),\qquad T_{\text{H}}\equiv \frac{\kappa_{e}}{2\pi}=\frac{f^{\prime}(r_{e})/2}{2\pi}=\frac{r_{c}^2+r_cr_e-2r_e^2}{4\pi r_{e}(r_c^2+r_cr_e+r_e^2)}\, .
\end{eqnarray}

\begin{figure}[htbp]
    \centering
    \subfigure[]
    {\includegraphics[width=0.45\linewidth]{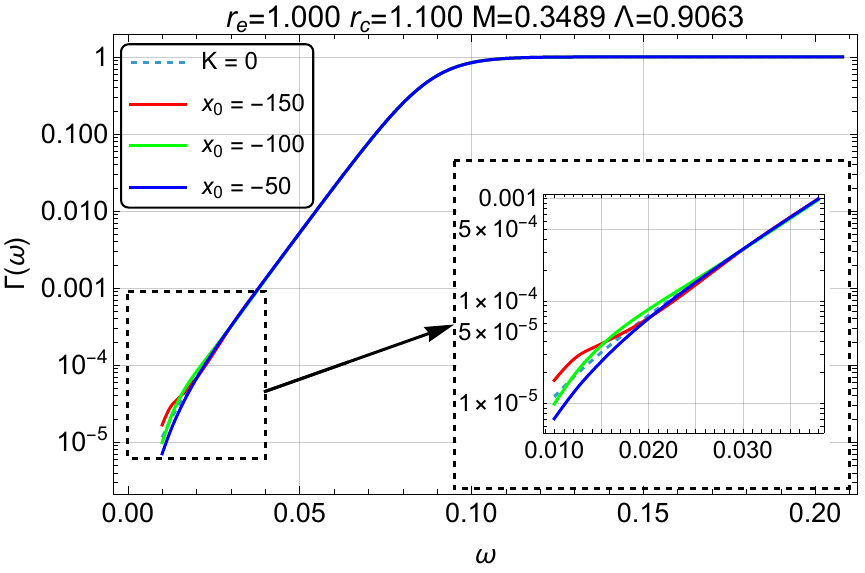}}\hfill
    \subfigure[]
     {\includegraphics[width=0.45\linewidth]{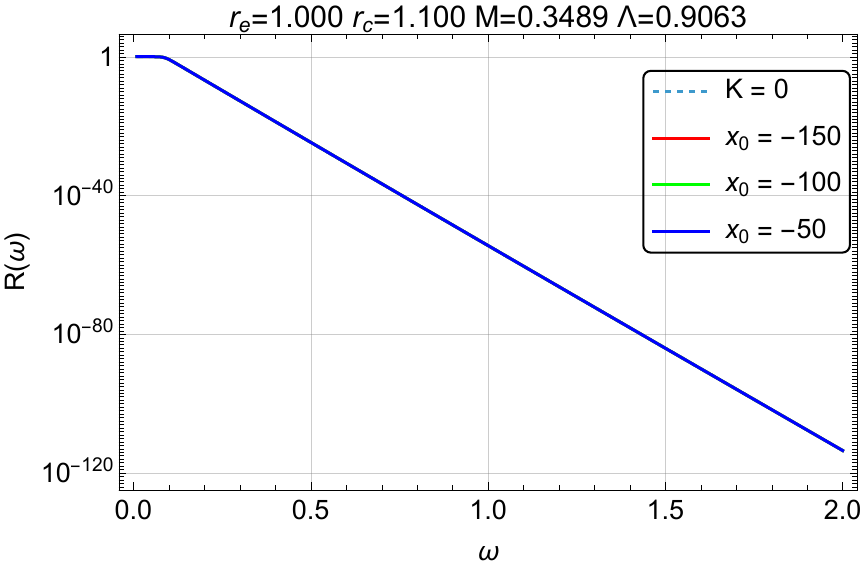}}
    \subfigure[]
    {\includegraphics[width=0.45\linewidth]{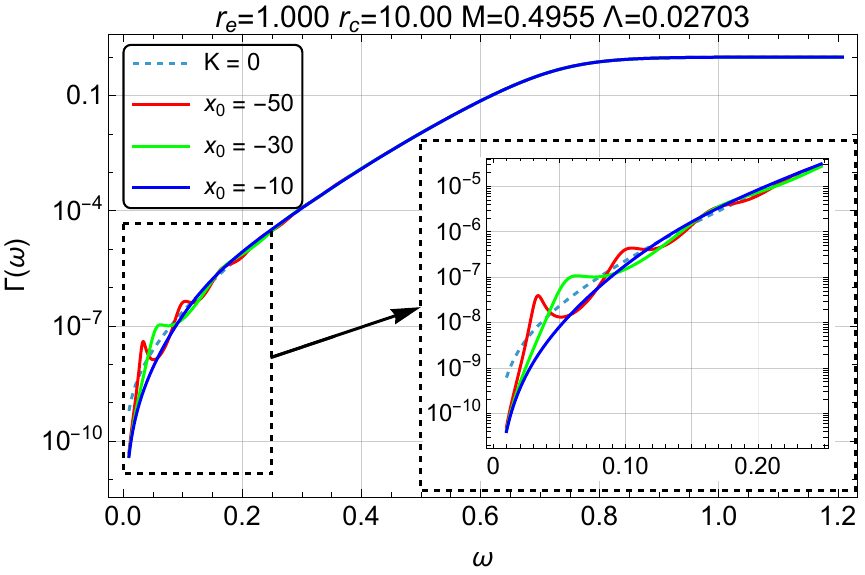}}\hfill
    \subfigure[]
    {\includegraphics[width=0.45\linewidth]{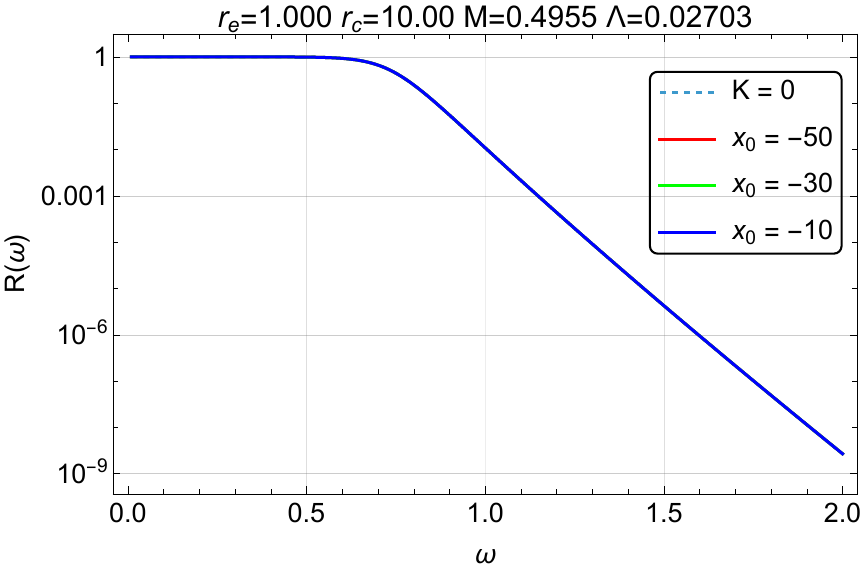}}
    \subfigure[]
    {\includegraphics[width=0.45\linewidth]{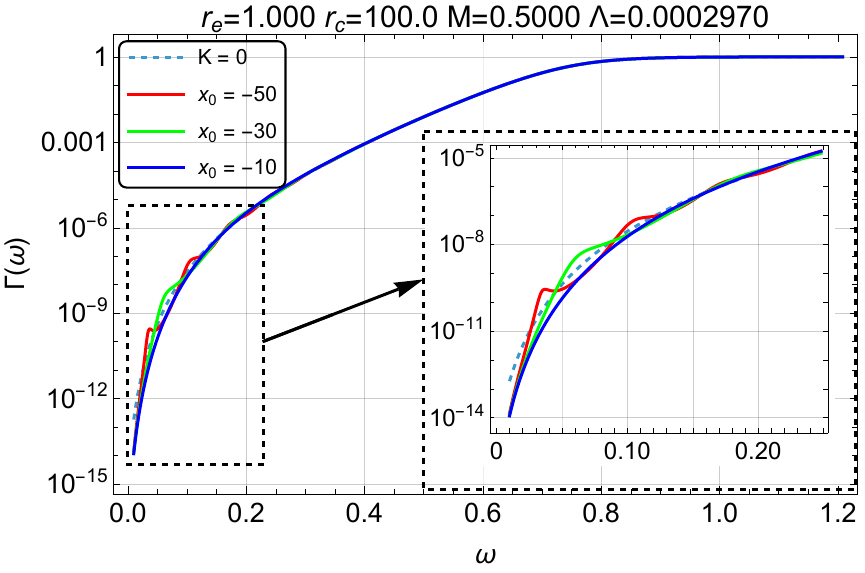}}\hfill
    \subfigure[]
    {\includegraphics[width=0.45\linewidth]{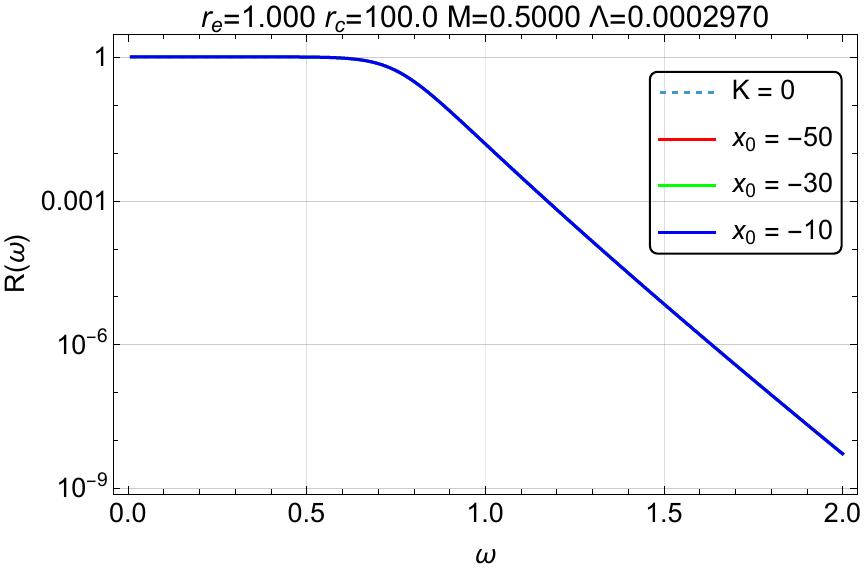}}\hfill
    \caption{Greybody factors $\Gamma(\omega)$ and reflection coefficients $R(\omega)$ for axial gravitational $(s=2)$ perturbations of the Schwarzschild-de Sitter black hole in the Boltzmann-type reflectivity model. Panels (a), (c), and (e) show $\Gamma(\omega)$, whereas panels (b), (d), and (f) display the corresponding reflectivity $R(\omega)$ for the same parameters. In each panel, the curve labelled $\mathcal{K}=0$ corresponds to the greybody factor of the original non-reflective SdS black hole. The top row corresponds to the near-extremal case with $r_e=1$ and $r_c=1.1$, with the reflective wall positions $x_0=-150$, $-100$, and $-50$. The middle row corresponds to $r_e=1$ and $r_c=10$ with $x_0=-50$, $-30$, and $-10$, and the bottom row to $r_e=1$ and $r_c=100$ with the same set of $x_0$.}
    \label{GFsBoltzmann}
\end{figure}

From Eq. (\ref{eq:GF_K_Boltz}) and Eq. (\ref{GFs}), we show the corresponding greybody factors in Fig. \ref{GFsBoltzmann}, where the right panels are the reflection coefficients. In sharp contrast with the constant reflectivity case, where changing the reflectivity $\mathcal{K}$ or the wall position $x_0$ produces strong oscillations, here the results obtained from the Boltzmann reflectivity are close to the one with $\mathcal{K}=0$. The differences are so small that they only become visible when the results are plotted on a logarithmic scale and the low-frequency region is locally magnified, as in the insets of Fig. \ref{GFsBoltzmann}. For each left panels of Fig. \ref{GFsBoltzmann}, it is found that changing the position of the reflective surface leads to a corresponding shift in the locations of the oscillation peaks of the greybody factors. This behaviour is similar to the constant reflectivity case, but for the Boltzmann reflectivity scenario the effect is much less evident. Finally, we comment on the behaviour of the reflection coefficients $R(\omega)$ in the high-frequency regime. As $\omega \gg 1$, $\mathcal{K}(\omega)$ is exponentially suppressed then we have
\begin{eqnarray}
    \lim_{\omega\to\infty} R(\omega)
    = \Bigg|\frac{A_{\text{out}}(\omega)}{A_{\text{in}}(\omega)}\Bigg|^2\, .
\end{eqnarray}
Thus, the high-frequency part of $R(\omega)$ for the Boltzmann case is simply the standard reflection coefficient of the non-reflective case. The straight, sloping line in the logarithmic plot of Fig. \ref{GFsBoltzmann} therefore just reflects the asymptotic decay of $R(\omega)$, while any contribution from the reflective wall becomes completely negligible in high-frequency regime.

\section{Exceptional points}\label{exceptional_points}
A widely accepted conclusion is that generating a second-order EP (two-fold degeneracy of eigenvalues) typically requires two free real parameters~\cite{Yang:2025dbn}. When there are insufficient free parameters, the phenomenon of EPs generally does not occur, and it is instead replaced by mode repulsion or avoided crossing~\cite{Berti:2025hly,Dias:2021yju,Motohashi:2024fwt,Oshita:2025ibu,Lo:2025njp,Takahashi:2025uwo}. For example, considering the QNM spectra of the Kerr black hole, there is only one dimensionless parameter, i.e., $a/M$. For $0.875\le a/M\le 0.915$, the mode $\omega_{225}$ resonates with the mode $\omega_{226}$, and for $0.95\le a/M\le1-10^{-6}$, the mode $\omega_{315}$ resonates with the mode $\omega_{316}$~\cite{Motohashi:2024fwt}. The physical quantity used to quantitatively characterize resonance phenomena is called the excitation factor (EF), which has been widely studied in many gravitational systems~\cite{Leaver:1986gd,Sun:1988tz,Andersson:1995zk,Glampedakis:2003dn,Berti:2006wq,Zhang:2013ksa,Silva:2024ffz,Oshita:2024wgt,Oshita:2025ibu,Lo:2025njp,Takahashi:2025uwo,Oshita:2021iyn,Chen:2024hum,Motohashi:2024fwt,Kubota:2025hjk,Wu:2025sbq}. As such type of the repulsion phenomenon occurs, the excitation factor will become very large for some appropriate parameters. At the same time, either the real parts of the modes cross and the imaginary parts repel, or the imaginary parts of the modes cross and the real parts repel. Therefore, if only one parameter is varied, the avoided crossing occurs.

However, when the number of free (real) parameters becomes $2$, the story changes. For example, in~\cite{Cavalcante:2024kmy,Cavalcante:2024swt}, a scalar field of mass $\mu$ around a Kerr black hole is considered. For $(M\mu)_c\simeq0.3704981$ and $(a/M)_c\simeq0.9994660$, it is found that the longest-living mode and the first overtone coincide, and this pair of parameters is called the EP. The hysteresis phenomenon of spectra also occurs, which is interpreted through a thermodynamic analogy. One can refer more details to the recent work~\cite{Cavalcante:2025abr}. Returning to our study, in Sec. \ref{QNMs}, we have obtained the QNM spectra migration due to the reflectivity $\mathcal{K}$ varies in the interval $[0,1]$. Therein, given the position of the wall $x_0$, and the cosmological constant via $r_c$ with the unit being $r_e=1$, the only parameter that can be freely adjusted is $\mathcal{K}\in\mathbb{R}$. So, as expected, there is no EP as $\mathcal{K}$ varies (at least for $\mathcal{K}\in[0,1]$). However, in this section, we allow the reflectivity $\mathcal{K}$ to be a complex constant, $\mathcal{K}\in\mathbb{C}$, which can provide two free real parameters. An imaginary part of the reflectivity denotes that the reflection process around the event horizon imparts a phase change of the wave, as opposed to a pure change in intensity. Doing so would lead to the appearance of exceptional points. This is also the first implementation of EPs caused by changes in QNM boundary conditions in the gravity aspect.

\begin{figure}[htbp]
    \centering
    \includegraphics[width=0.55\linewidth]{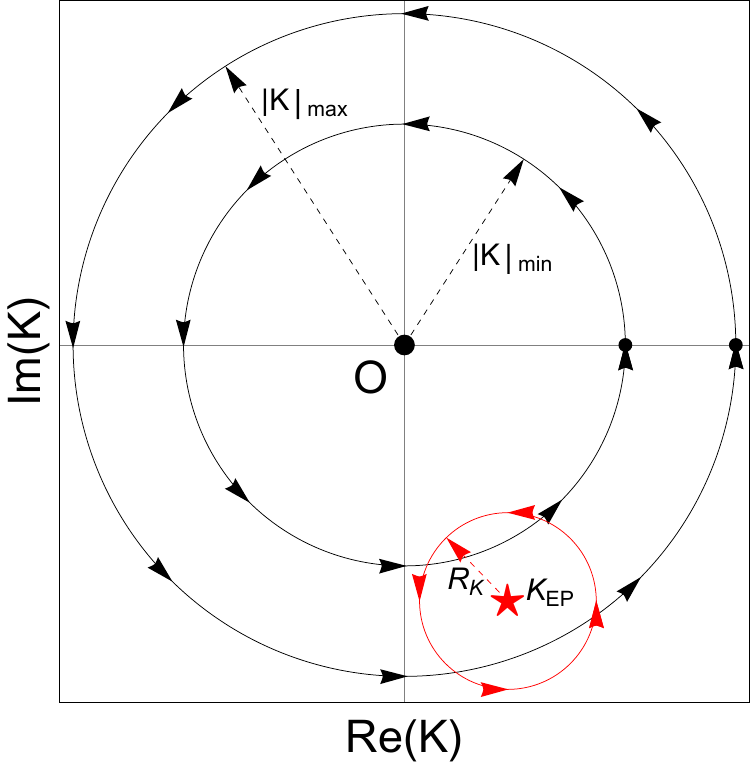}
    \caption{The schematic diagram of the EP finding method, where the red star stands for the EP denoting $\mathcal{K}_{\text{EP}}$. The radius of a large circle is $|\mathcal{K}|_{\text{min}}$, and the radius of a small circle is $|\mathcal{K}|_{\text{max}}$. In practice, the difference between $|\mathcal{K}|_{\text{min}}$ and $|\mathcal{K}|_{\text{max}}$ is very small. Two black points on the positive real half axis are the start points. Here, we have $\text{Arg}(\mathcal{K}_{\text{EP}})\simeq-67.9959^{\circ}$. Red circle corresponds to Eq. (\ref{small_circumference}).}
    \label{EP_Finding_schematic_diagram}
\end{figure}

\begin{figure}[htbp]
    \centering
    \subfigure[]
     {\includegraphics[width=0.46\linewidth]{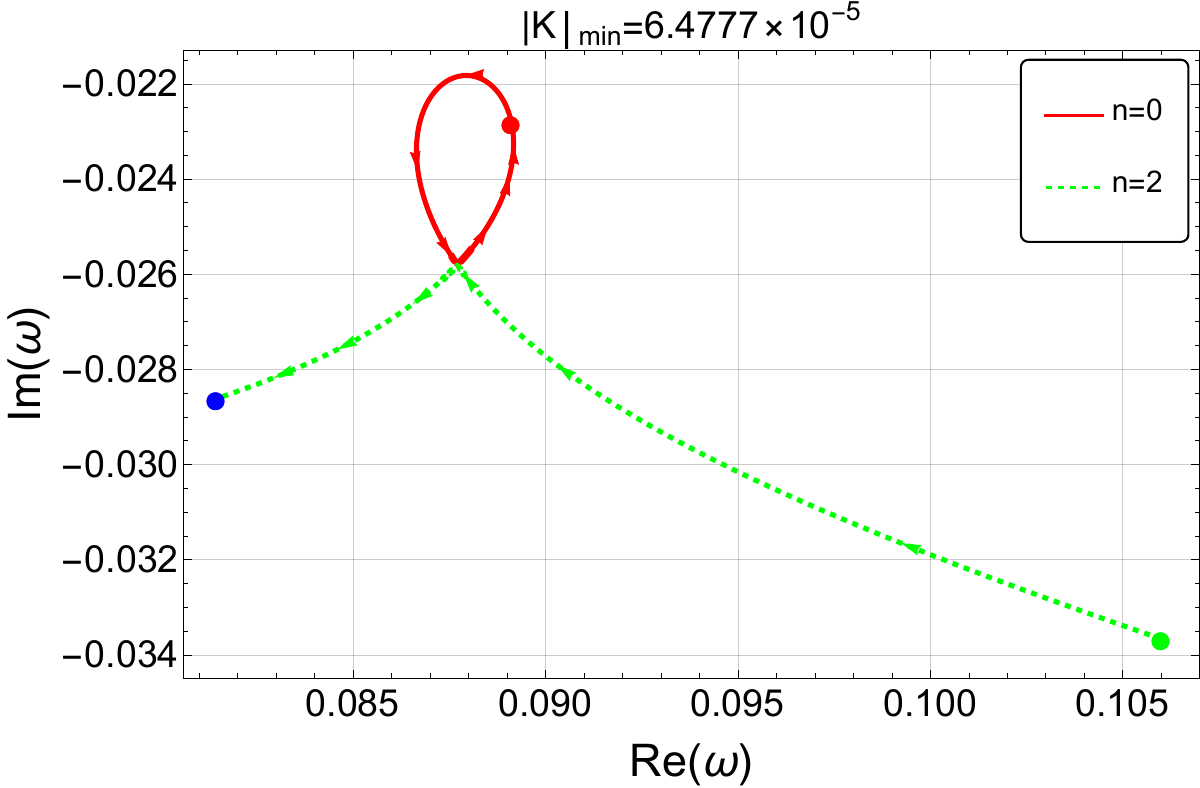}}\hfill
     \subfigure[]
    {\includegraphics[width=0.46\linewidth]{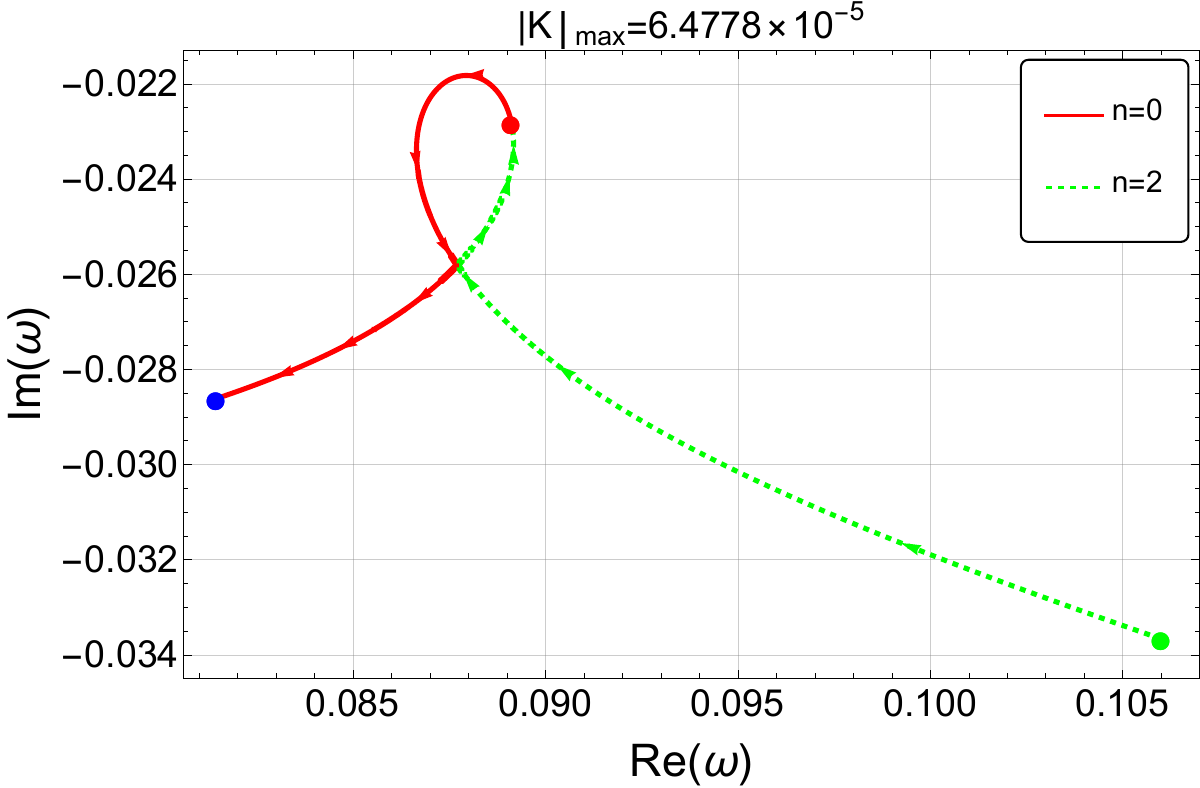}}
    \caption{The migrations for fundamental mode $n=0$ (red solid line) and the overtone $n=2$ (green dashed line) within the parameter $r_e=1$, $r_c=1.1$, $x_0=-150$, where panel (a) corresponds to the parameter circumference $|\mathcal{K}|=|\mathcal{K}|_{\text{min}}$ while panel (b) corresponds to the parameter circumference $|\mathcal{K}|=|\mathcal{K}|_{\text{max}}$. The red point, blue point and green point represent the positions for the original fundamental mode $n=0$, the original first overtone $n=1$ and the original second overtone $n=2$, respectively.}
    \label{EP_Finding}
\end{figure}

Now, we will search for the EP of $\mathcal{K}$ on the complex plane. The QNM sepctra results for the real $\mathcal{K}>0$ are cornerstones of our EP finding approach. The schematic diagram for the EP finding approach is given in Fig. \ref{EP_Finding_schematic_diagram}. On the one hand, for a small $|\mathcal{K}|_{\text{min}}$, one starts from the left black point in Fig. \ref{EP_Finding_schematic_diagram}, in order to implement the outcomes of the QNM spectra on the circumference $|\mathcal{K}|=|\mathcal{K}|_{\text{min}}$. On the other hand, for a large $|\mathcal{K}|_{\text{max}}$, one starts from the right black point in Fig. \ref{EP_Finding_schematic_diagram}, in order to implement the outcomes of the QNM spectra on the circumference $|\mathcal{K}|=|\mathcal{K}|_{\text{max}}$. Continuously narrow the gap between $|\mathcal{K}|_{\text{min}}$ and $|\mathcal{K}|_{\text{max}}$ so that the mode repulsion phenomenon shown in Fig. \ref{EP_Finding} occurs, in which we have chosen typical parameters. From the left panel of Fig. \ref{EP_Finding}, as the complex parameter $\mathcal{K}$ move along the circumference $|\mathcal{K}|_{\text{min}}$, the fundamental mode $n=0$ goes back the original position, and the overtone $n=2$ migrates to the position for the original overtone $n=1$. From the right panel of Fig. \ref{EP_Finding}, as the complex parameter $\mathcal{K}$ moves along the circumference $|\mathcal{K}|_{\text{max}}$, the fundamental mode $n=0$ migrates to the position for the original overtone $n=1$, and the overtone $n=2$ migrates to the original position for the fundamental mode $n=0$. One can find a peculiar parameter $\mathcal{K}_{\text{EP}}$ so that the red solid line is closest to the green dashed line regardless of in the left panel or in the right panel of Fig. \ref{EP_Finding}. Such point, which is depicted as a red star, is nothing but the exceptional point. For the parameter in Fig. \ref{EP_Finding}, the EP for the fundamental mode and the second overtone is given by $\mathcal{K}_{\text{EP}}\simeq2.42704\times10^{-5}-6.00589\times10^{-5}\mathrm{i}$. It is worth noting that in the process of moving the parameter, no exchange between two modes $n=0$ and $n=2$ actually occurred. This phenomenon arises because the loops, $|\mathcal{K}|=|\mathcal{K}|_{\text{min}}$ and $|\mathcal{K}|=|\mathcal{K}|_{\text{max}}$, actually contains infinitely many EPs~\cite{Cavalcante:2025abr}, rather than just one. Some of these EPs involves the $n=0$ mode or $n=2$ mode coalescence with another different modes. The method given above can actually be considered a relatively reliable way to find EPs, especially for cases where the equations of QNM spectra can be analytically established.

\begin{figure}[htbp]
    \centering
    \subfigure[]
     {\includegraphics[width=0.46\linewidth]{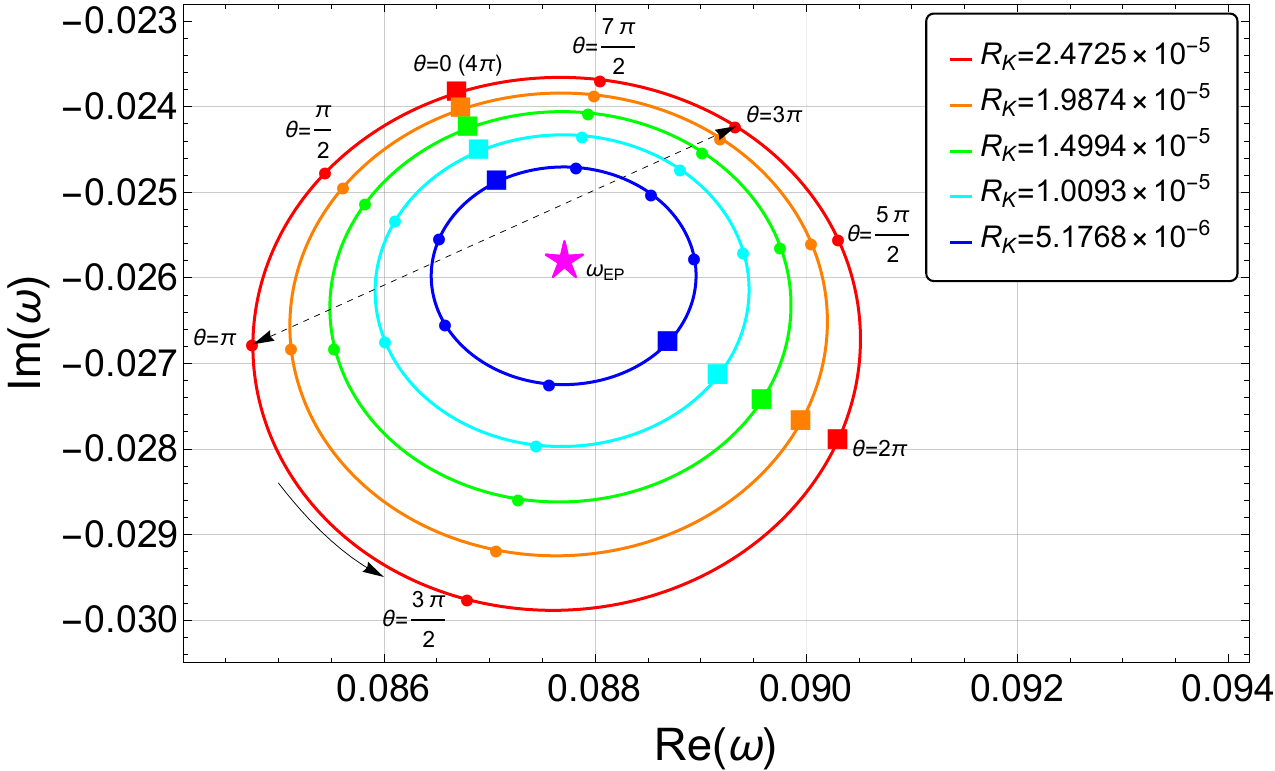}\label{EP_plot_EPs}}\hfill
     \subfigure[]
    {\includegraphics[width=0.46\linewidth]{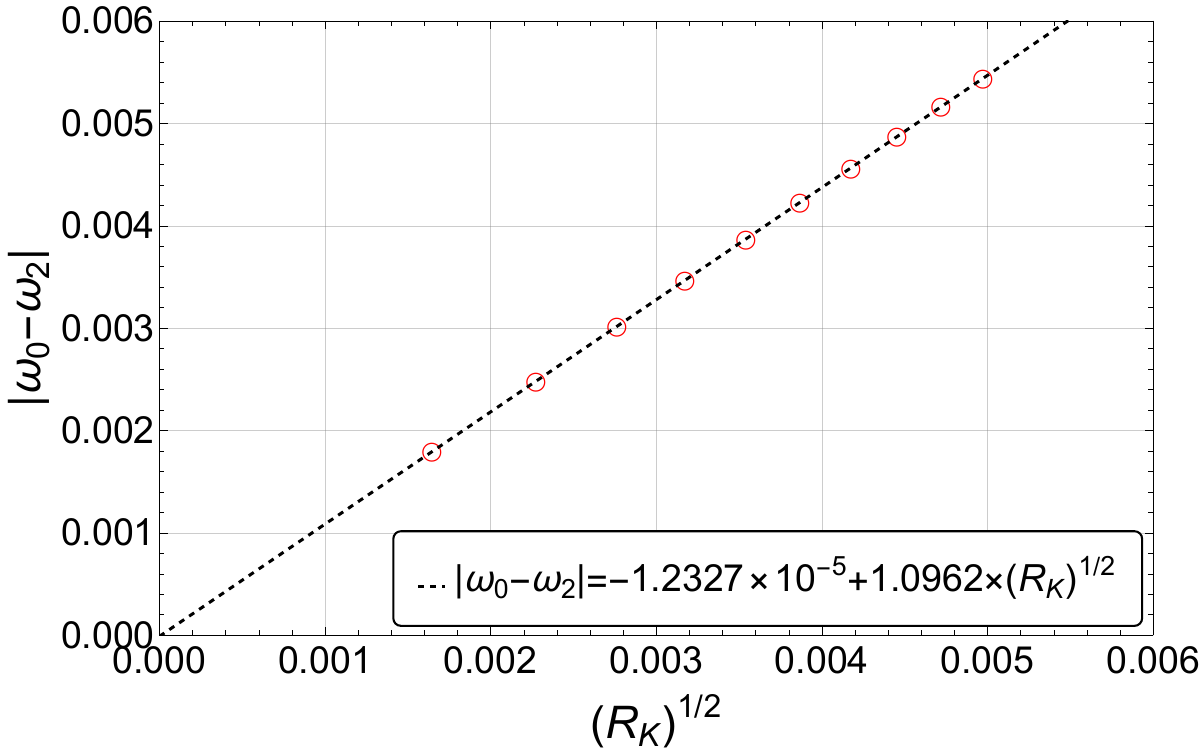}\label{EP_plot_Diff_omega_RK_Fit_Plot}}
    \caption{Panel (a) shows the migrations for fundamental mode $n=0$ with the parameter $r_e=1$, $r_c=1.1$, $x_0=-150$, for the parametric equation (\ref{small_circumference}) of the small circumference. Different color lines correspond to the different radii. On each lines, the symbol $\blacksquare$ corresponds to $\theta=0$ and $\theta=2\pi$, the symbol $\bullet$ corresponds to $\theta=\pi/2$, $\pi$, $3\pi/2$, $5\pi/2$, $3\pi$ and $7\pi/2$ and $\theta=2\pi$. The magenta star is the degenerate QNM spectra for the EP. In addition, the migration for second overtone starts from $\theta=2\pi$. $|\omega_0-\omega_2|$ v.s. $(R_{\mathcal{K}})^{1/2}\equiv|\mathcal{K}-\mathcal{K}_{\text{EP}}|^{1/2}$ is shown in panel (b), and the dashed line is corresponding linear fit.}
    \label{EP_plot}
\end{figure}

To avoid the influence of other EPs, we consider a small circumference around a certain EP (see the small red circle in Fig. \ref{EP_Finding_schematic_diagram}). The parametric equation for such small circumference is given by
\begin{eqnarray}\label{small_circumference}
    \mathcal{K}=\mathcal{K}_{\text{EP}}+R_{\mathcal{K}}\cdot\mathrm{e}^{\mathrm{i}\theta}\, ,
\end{eqnarray}
where $R_{\mathcal{K}}$ is the radius of the small circumference, and $\theta$ is the angle parameter. The parameter range is $[0,4\pi]$, which is equivalent to the parameter circling around the EP twice. We take parameters $r_e=1$, $r_c=1.1$ and $x_0=-150$ as an example to illustrate, where one of the EPs is $\mathcal{K}_{\text{EP}}\simeq2.42704\times10^{-5}-6.00589\times10^{-5}\mathrm{i}$, and other parameters are also treated similarly. In left panel of Fig. \ref{EP_plot}, the migrations of the fundamental modes are shown, in which different colors represent different radii. In fact, the results of the left panel of Fig. \ref{EP_plot} is also the migration of the second overtone, where the difference is that the mode $n=2$ starts from $\theta=2\pi$ and ends at $\theta=2\pi$. The left of the Fig. \ref{EP_plot} shows the exchange of modes, which is known as the so-called the hysteresis phenomenon of QNMs~\cite{Cavalcante:2024kmy,Cavalcante:2024swt,Cavalcante:2025abr,Motohashi:2024fwt}. Furthermore, regarding black hole spectroscopy, we will demonstrate that the necessity of applying Puiseux series near the EP in the parameterized analysis of QNMs~\cite{Cao:2025afs}. In order to show it, for a given $R_{\mathcal{K}}$, averaging the distance between $\omega_0$ and $\omega_2$ with $\theta\in[0,4\pi]$, where the dashed double arrow line represents this distance depicted in Fig. \ref{EP_plot_EPs}. The derived average gives the quantity which is denoted as $|\omega_0-\omega_2|$. It can be plotted that the relation between $|\omega_0-\omega_2|$ and $(R_{\mathcal{K}})^{1/2}$ in Fig. \ref{EP_plot_Diff_omega_RK_Fit_Plot}. From the Fig. \ref{EP_plot_Diff_omega_RK_Fit_Plot}, we fit the numerical results of $|\omega_0-\omega_2|$ and $R_{\mathcal{K}}$ to see that $|\omega_0-\omega_2|$ is proportional to $R_{\mathcal{K}}^{1/2}$, giving the structure of the Puiseux series at around the EP~\cite{Ashida:2020dkc} [cf. Eq. (111) therein].

\section{Conclusions and discussion}\label{conclusions}
This work investigates the perturbation of Schwarzschild-de Sitter black holes in semi-open system, which is formulated by introducing a partially reflective wall near the event horizon. This setup aims to model realistic scenarios where black holes are not perfect absorbers such as exotic compact objects. Utilizing the mathematical framework of Heun functions, the study provides an exact analytical treatment of the perturbation equations governing scalar, electromagnetic, and axial gravitational perturbations. We systematically explore how the introduction of reflectivity influences three key characteristics of black holes, namely the QNM spectra, the greybody factors, and the possible emergence of exceptional points.

In the theoretical setup, we begin by reviewing the geometry of the SdS spacetime and cast the perturbation equation into the form of Heun's equation (cf. Appendix \ref{Heun_functions}). Local analytic solutions near the event and cosmological horizons are obtained (see Eq. (\ref{general_solution_z_0}) and Eq. (\ref{general_solution_z_1})). Connection coefficients between solutions in different regions are established using Wronskians. The boundary conditions for present semi-open system are precisely defined as: at a fixed position near the event horizon, the wave function is a superposition of ingoing and outgoing waves, and at the cosmological horizon, only outgoing waves are required. Hence, the resonance condition determining the QNM spectra is derived, which is written as $S_{\text{in}}(\omega)=0$. 

Regarding the QNM spectra (see Sec. \ref{QNMs}), we focuses on axial gravitational perturbations $(s=2)$ and examine the impact of a real, constant reflectivity, i.e., $\mathcal{K}\in[0,1]$. Two paradigmatic cases are compared, the near-extremal SdS black hole and the near Schwarzschild case. Numerical computations reveal that as the reflectivity increases from $0$ (open system) to $1$ (perfectly reflective wall), the QNMs undergo systematic migration in the complex plane. Three distinct types of migrations are identified: a subset of modes approaches the real axis, forming long-lived QBS; another subset moves toward but does not reach the real axis retaining a non-zero decay rate; the third type, primarily higher overtones, veers toward and eventually lands on the imaginary axis, becoming purely decaying modes. It is also found that the position of the reflective wall affects spectrum stability. To be more specific, walls placed closer to the event horizon render the spectrum more sensitive to small changes in $\mathcal{K}$, indicating enhanced instability.

Concerning the greybody factors (see Sec. \ref{graybody_factors}), the influence of reflectivity on them is studied. For constant reflectivity, for $\mathcal{K}>0$, pronounced resonant oscillation peaks appear in the curves, corresponding to resonant modes within the effective cavity formed between the reflective wall and the potential. The number and spacing of these resonance peaks depends on the ``effective length'' between the wall and the potential. A larger $|x_0|$ results in denser peaks. For comparison, the study also examines a physically motivated Boltzmann-type reflectivity $\mathcal{K}(\omega)=\exp(-|\omega|/T_{\text{H}})$. For this model, the derived GF differs only minimally from the standard black hole case, with slight differences at low frequencies.

Regarding the exceptional point (see Sec. \ref{exceptional_points}), the present study achieves that in the context of gravitational perturbations by extending the reflectivity to a complex constant thereby successfully constructing and identifying the (second-order) EP. By parameterizing on a small circle around the EP and varying the parameter, a mode exchange phenomenon is observed (see Fig. \ref{EP_plot_EPs}). The identities of the two modes swap after encircling the EP, confirming the hysteretic nature of the spectrum. Further analysis shows that near the EP, the frequency difference scales proportionally to the square root of the parameter deviation. This scaling is consistent with the Puiseux series expansion expected near an EP. In Fig. \ref{EP_plot_Diff_omega_RK_Fit_Plot}, we show the numerical verification of the theoretical prediction.

In conclusion, this work employs a combined analytic (Heun function) and numerical approach to comprehensively elucidate the impact of semi-open systems on the perturbation of SdS black holes. The extension of this work is concentrating on the Kerr-de Sitter black holes in semi-open systems~\cite{Hatsuda:2020sbn}. For this rotation context, exceptional points and lines~\cite{Cao:2025afs} caused by boundary conditions can be studied, and even higher-order exceptional points could be found. The ringdown waveform corresponding to EPs caused by boundary conditions can be further studied, similar to~\cite{Yang:2025dbn,PanossoMacedo:2025xnf}.

\section*{Acknowledgement}
This work is supported by the National Natural Science Foundation of China with grant No. 12505067. This work is also supported in part by the National Key R\&D Program of China Grant No. 2022YFC2204603, by the National Natural Science Foundation of China with grants No.12475063, No. 12075232.

\appendix
\section{The Heun's equation}\label{Heun_functions}
In the appendix, we will solve Eq. (\ref{master_equation}) by using the Heun function, which is a generalization of the hypergeometric function. It is not difficult to see that the differential equation (\ref{master_equation}) has four regular singular points at $r=r_n$, $r=0$, $r=r_e$, $r=r_c$. Note that the point $r=\infty$ is the regular point for all cases. In terms of $r$, Eq. (\ref{master_equation}) becomes a standard form
\begin{eqnarray}\label{standard_equation}
    \frac{\mathrm{d}^2\Psi}{\mathrm{d}r^2}+p(r)\frac{\mathrm{d}\Psi}{\mathrm{d}r}+q(r)\Psi=0\, ,
\end{eqnarray}
where the functions $p(r)$ and $q(r)$ read
\begin{eqnarray}
    p(r)=\frac{f^{\prime}(r)}{f(r)}\, ,\quad q(r)=\frac{\omega^2-V_s(r)}{f^2(r)}\, . 
\end{eqnarray}
We calculate the indices at each singular points, where the indices are denoted by $\rho_n$, $\rho_0$, $\rho_e$, $\rho_c$, respectively. The results are summaried in the Tab. \ref{indices}.

\begin{table}[t]
\setlength{\tabcolsep}{12pt} 
\centering
\begin{tabular}{c c c c c}
\hline\hline
Singular point & Spin $(s)$ & $a_0$ & $b_0$ & Indices $(\rho_{1},\rho_{2})$ \\
\hline\noalign{\vskip 4pt}
$r=r_n$ & any & $1$ & $\displaystyle \frac{\omega^2}{[f'(r_n)]^2}$ & $\displaystyle \Bigl(+\frac{\mathrm{i}\omega}{f^{\prime}(r_n)},\,-\frac{\mathrm{i}\omega}{f^{\prime}(r_n)}\Bigr)$ \\[4pt]
\noalign{\vskip 4pt}\hline\noalign{\vskip 4pt}
\multirow{3}{*}{$r = 0$} 
& $0$ & $-1$ & $1$ & $(1,\,1)$ \\[2pt]
& $1$ & $-1$ & $0$ & $(2,\,0)$ \\[2pt]
& $2$ & $-1$ & $-3$ & $(3,\,-1)$ \\[2pt]
\hline\noalign{\vskip 4pt}
$r=r_e$& any & $1$ & $\displaystyle \frac{\omega^2}{[f^{\prime}(r_e)]^2}$ & $\displaystyle \Bigl(+\frac{\mathrm{i}\omega}{f^{\prime}(r_e)},\,-\frac{\mathrm{i}\omega}{f^{\prime}(r_e)}\Bigr)$ \\[6pt]
\noalign{\vskip 4pt}\hline\noalign{\vskip 4pt}
$r=r_c$& any & $1$ & $\displaystyle \frac{\omega^2}{[f^{\prime}(r_c)]^2}$ & $\displaystyle \Bigl(+\frac{\mathrm{i}\omega}{f^{\prime}(r_c)},\,-\frac{\mathrm{i}\omega}{f^{\prime}(r_c)}\Bigr)$ \\
\noalign{\vskip 4pt}\hline\hline
\end{tabular}
\caption{Singular points of Eq. (\ref{standard_equation}), the corresponding coefficients $a_0,b_0$ in the index equation $\rho(\rho-1)+a_0\rho+b_0=0$, and the associated indices $(\rho_1,\rho_2)$ for scalar $(s=0)$, electromagnetic $(s=1)$, and axial gravitational $(s=2)$ perturbations.}
\label{indices}
\end{table}

We consider the M\"{o}bius transformation as follow
\begin{eqnarray}\label{Mobius_transformation}
    z=\frac{r_c(r-r_e)}{(r_c-r_e)r}\, .
\end{eqnarray}
In this transformation, $r=r_n$, $0$, $r_e$, $r_c$ are mapped into $z_n$, $\infty$, $0$, $1$, respectively, where
\begin{eqnarray}
    z_n=\frac{r_c(r_n-r_e)}{(r_c-r_e)r_n}>1\, .
\end{eqnarray}
These four points happen to be the four regular singular points of the Heun's equation. We next perform the scaling transformation of $\Psi(z)$:
\begin{eqnarray}
    \Psi(z)=z^{\rho_{e,1}}(z-1)^{\rho_{c,1}}(z-z_n)^{\rho_{n,1}}u(z)\, .
\end{eqnarray}
The new function $u(z)$ will satisfy the Heun's equation
\begin{eqnarray}\label{Heun_equation}
    \frac{\mathrm{d}^2u(z)}{\mathrm{d}z^2}+\Big(\frac{\gamma}{z}+\frac{\delta}{z-1}+\frac{\epsilon}{z-z_n}\Big)\frac{\mathrm{d}u(z)}{\mathrm{d}z}+\frac{(\alpha\beta z-q)u(z)}{z(z-1)(z-z_n)}=0\, ,
\end{eqnarray}
where $\gamma+\delta+\epsilon=\alpha+\beta+1$. One can apply the Frobenius method to get the infinite power series solutions. In particular, two local solutions at $z=0$ are given by~\cite{Oshita:2022pyf,Hatsuda:2020sbn,Oshita:2021iyn}
\begin{eqnarray}\label{u01_u02}
    u_{01}(z)&=&Hl(z_n,q;\alpha,\beta,\gamma,\delta;z)\, ,\nonumber\\
    u_{02}(z)&=&z^{1-\gamma}Hl(z_n,(z_n\delta+\epsilon)(1-\gamma)+q;\alpha+1-\gamma,\beta+1-\gamma,2-\gamma,\delta;z)\, ,
\end{eqnarray}
while two local solutions at $z=1$ are given by
\begin{eqnarray}\label{u11_u12}
    u_{11}(z)&=&Hl(1-z_n,\alpha\beta-q;\alpha,\beta,\delta,\gamma;1-z)\, ,\nonumber\\
    u_{12}(z)&=&(1-z)^{1-\delta}Hl(1-z_n,((1-z_n)\gamma+\epsilon)(1-\delta)+\alpha\beta-q;\alpha+1-\delta,\beta+1-\delta,2-\delta,\gamma;1-z)\, .
\end{eqnarray}

Next, we give parameters in the Heun's equation for all cases, where the condition $r_n=-(r_e+r_c)$ has been used. Note that $\alpha$ and $\beta$ can be exchanged. For the three types of perturbations, they share the same $\gamma$, $\delta$, and $\epsilon$, namely,
\begin{eqnarray}\label{gamma}
    \gamma(\omega)=-\frac{\Lambda r_c^3+\Lambda r_c^2 r_e+r_cr_e(-2\Lambda r_e+6\mathrm{i}\omega)}{\Lambda (r_c-r_e)(-r_c^2-2r_c r_e)}\, ,
\end{eqnarray}
\begin{eqnarray}\label{delta}
    \delta(\omega)=\frac{-2\Lambda r_c^2 r_e+r_c r_e (\Lambda  r_e+6 \mathrm{i} \omega)+\Lambda  r_e^3}{\Lambda  (r_c-r_e) (-2 r_c r_e-r_e^2)}\, ,
\end{eqnarray}
and
\begin{eqnarray}\label{epsilon}
    \epsilon(\omega)=\frac{2 \Lambda  r_c^2+5 \Lambda r_c r_e+6 \mathrm{i} r_c \omega +2 \Lambda  r_e^2+6 \mathrm{i} r_e \omega}{2 \Lambda  r_c^2+5 \Lambda r_c r_e+2 \Lambda  r_e^2}\, ,
\end{eqnarray}
where $\Lambda$ is given by Eqs. (\ref{mass_and_cosmological_constant}). For $s=0$, the remaining parameters are
\begin{eqnarray}
    q=\frac{-3 \ell^2-3 \ell-\Lambda r_c (r_c+r_e)}{\Lambda  (-r_c-r_e) (r_c-r_e)}\, ,\quad \alpha=1\, ,\quad \beta=1\, .
\end{eqnarray}
For $s=1$, the remaining parameters are
\begin{eqnarray}
    q=-\frac{3 \ell (\ell+1)}{\Lambda (-r_c-r_e)(r_c-r_e)}\, ,\quad \alpha=0\, ,\quad \beta=2\, .
\end{eqnarray}
For $s=2$, the remaining parameters are
\begin{eqnarray}
    q=-\frac{3\Big[\ell^2+\ell-\Lambda r_c (r_c+r_e)\Big]}{\Lambda  (-r_c-r_e) (r_c-r_e)}\, ,\quad  \alpha=-1\, ,\quad \beta=3\, .
\end{eqnarray}

\section{The expressions of four functions for $A_{\text{in}}(\omega)$, $A_{\text{out}}(\omega)$, $B_{\text{in}}(\omega)$ and $B_{\text{out}}(\omega)$}\label{expressions_Ain_Aout_Bin_Bout}
In this appendix, we give the detailed process of getting four functions for $A_{\text{in}}(\omega)$, $A_{\text{out}}(\omega)$, $B_{\text{in}}(\omega)$ and $B_{\text{out}}(\omega)$. It is not hard to find that the solution $\Psi_{11}$ is proportional to the ``up'' solution $\Psi_{\text{up}}$. Therefore, one can use the solution $\Psi_{11}$ to solve $B_{\text{in}}(\omega)$ and $B_{\text{out}}(\omega)$. From the definition of the tortoise coordinate (\ref{tortoise_coordinates_integration}) and the M\"{o}bius transformation (\ref{Mobius_transformation}), we have the relations between $\{\mathrm{e}^{\mathrm{i}\omega x}\, ,\mathrm{e}^{-\mathrm{i}\omega x}\}$ and $z$, i.e.,
\begin{eqnarray}
    \mathrm{e}^{\mathrm{i}\omega x}&=&\Big(1-\frac{r_e}{r_c}\Big)^{\rho_{e,1}}\Big(\frac{r_c}{r_e}-1\Big)^{\rho_{c,1}}\Big(\frac{r_n}{r_e}-\frac{r_n}{r_c}\Big)^{\rho_{n,1}}z^{\rho_{e,1}}(1-z)^{\rho_{c,1}}(z-z_n)^{\rho_{n,1}}\, ,\label{useful_relations_1}\\
    \mathrm{e}^{-\mathrm{i}\omega x}&=&\Big(1-\frac{r_e}{r_c}\Big)^{\rho_{e,2}}\Big(\frac{r_c}{r_e}-1\Big)^{\rho_{c,2}}\Big(\frac{r_n}{r_e}-\frac{r_n}{r_c}\Big)^{\rho_{n,2}}z^{\rho_{e,2}}(1-z)^{\rho_{c,2}}(z-z_n)^{\rho_{n,2}}\, .\label{useful_relations_2}
\end{eqnarray}
Using Eqs. (\ref{relations_u}) and Eqs. (\ref{u01_u02}), the solution $\Psi_{11}$ is expressed as
\begin{eqnarray}\label{Psi_11_Hl}
    \Psi_{11}&=&z^{\rho_{e,1}}(z-1)^{\rho_{c,1}}(z-z_n)^{\rho_{n,1}}\Big[C_{11}Hl(z_n,q;\alpha,\beta,\gamma,\delta;z)\nonumber\\
    &&+C_{12}z^{1-\gamma}Hl(z_n,(z_n\delta+\epsilon)(1-\gamma)+q;\alpha+1-\gamma,\beta+1-\gamma,2-\gamma,\delta;z)\Big]\, .
\end{eqnarray}
Combining Eq. (\ref{Psi_11_Hl}) with Eqs. (\ref{useful_relations_1}), one solves $B_{\text{out}}(\omega)$ and $B_{\text{in}}(\omega)$ which are given by
\begin{eqnarray}
    B_{\text{in}}(\omega)=C_{12}(\omega)/F(\omega)\, ,\quad B_{\text{out}}(\omega)=C_{11}(\omega)\, ,
\end{eqnarray}
where the function $F(\omega)$ is
\begin{eqnarray}
    F(\omega)= \Big(1-\frac{r_e}{r_c}\Big)^{1-\gamma(\omega)}\cdot\Big(\frac{r_c}{r_e}-1\Big)^{1-\delta(\omega)}\cdot\Big(1-\frac{r_n}{r_e}\Big)^{1-\epsilon(\omega)}\, .
\end{eqnarray}
The detail expressions of $\gamma(\omega)$, $\delta(\omega)$ and $\epsilon(\omega)$ are able to be found in Appendix \ref{Heun_functions}. It is also not hard to find that the solution $\Psi_{02}$ is proportional to the ``in'' solution $\Psi_{\text{in}}$ from which one can solve $A_{\text{in}}(\omega)$ and $A_{\text{out}}(\omega)$. Using Eqs. (\ref{relations_u_inverse}) and Eqs. (\ref{u11_u12}), the solution $\Psi_{02}$ is expressed as
\begin{eqnarray}\label{Psi_02_Hl}
    \Psi_{02}&=&z^{\rho_{e,1}}(z-1)^{\rho_{c,1}}(z-z_n)^{\rho_{n,1}}\Big[D_{21}Hl(1-z_n,\alpha\beta-q;\alpha,\beta,\delta,\gamma;1-z)\nonumber\\
    &&+D_{22}(1-z)^{1-\delta}Hl(1-z_n,((1-z_n)\gamma+\epsilon)(1-\delta)+\alpha\beta-q;\alpha+1-\delta,\beta+1-\delta,2-\delta,\gamma;1-z)\Big]\, .
\end{eqnarray}
One solves $A_{\text{in}}(\omega)$ and $A_{\text{out}}(\omega)$ that are expressed as
\begin{eqnarray}
    A_{\text{in}}(\omega)=G(\omega)\cdot D_{22}(\omega)\, ,\quad A_{\text{out}}(\omega)=F(\omega)\cdot D_{21}(\omega)\, ,
\end{eqnarray}
where the function $G(\omega)$ is
\begin{eqnarray}
    G(\omega)=\Big(1-\frac{r_n}{r_c}\Big)^{\epsilon(\omega)-1}\cdot\Big(1-\frac{r_n}{r_e}\Big)^{1-\epsilon(\omega)}\, .
\end{eqnarray}
It is difficult for us to prove that $A_{\text{in}}(\omega)=B_{\text{out}}(\omega)$ directly, but it can be checked numerically at least for general $\omega\in\mathbb{C}$. This means the reliability of our results. Go a step further, one writes down the analytical expression of the ``in'' solution and the ``up'' solution associated with $z$, namely
\begin{eqnarray}
    \Psi_{\text{in}}=Y_{\text{in}}(\omega)\cdot\Psi_{02}\, ,\quad \Psi_{\text{up}}=Y_{\text{up}}(\omega)\cdot\Psi_{11}\, ,
\end{eqnarray}
in which two factor functions $Y_{\text{in}}(\omega)$ and $Y_{\text{up}}(\omega)$ are
\begin{eqnarray}
    Y_{\text{in}}(\omega)=\Big(1-\frac{r_e}{r_c}\Big)^{\rho_{e,2}}\Big(\frac{r_c}{r_e}-1\Big)^{\rho_{c,2}}\Big(\frac{r_n}{r_e}-\frac{r_n}{r_c}\Big)^{\rho_{n,2}}(-1)^{\rho_{n,2}}\Big[\frac{r_c(r_e-r_n)}{(r_c-r_e)r_n}\Big]^{1-\epsilon(\omega)}\, ,
\end{eqnarray}
and
\begin{eqnarray}
    Y_{\text{up}}(\omega)=\Big(1-\frac{r_e}{r_c}\Big)^{\rho_{e,1}}\Big(\frac{r_c}{r_e}-1\Big)^{\rho_{c,1}}\Big(\frac{r_n}{r_e}-\frac{r_n}{r_c}\Big)^{\rho_{n,1}}(-1)^{\rho_{c,2}}\, .
\end{eqnarray}

\section{Quasinormal modes for $s=0$}\label{QNMs_s_0}
\begin{figure}[htbp]
    \centering
   \includegraphics[width=0.85\linewidth]{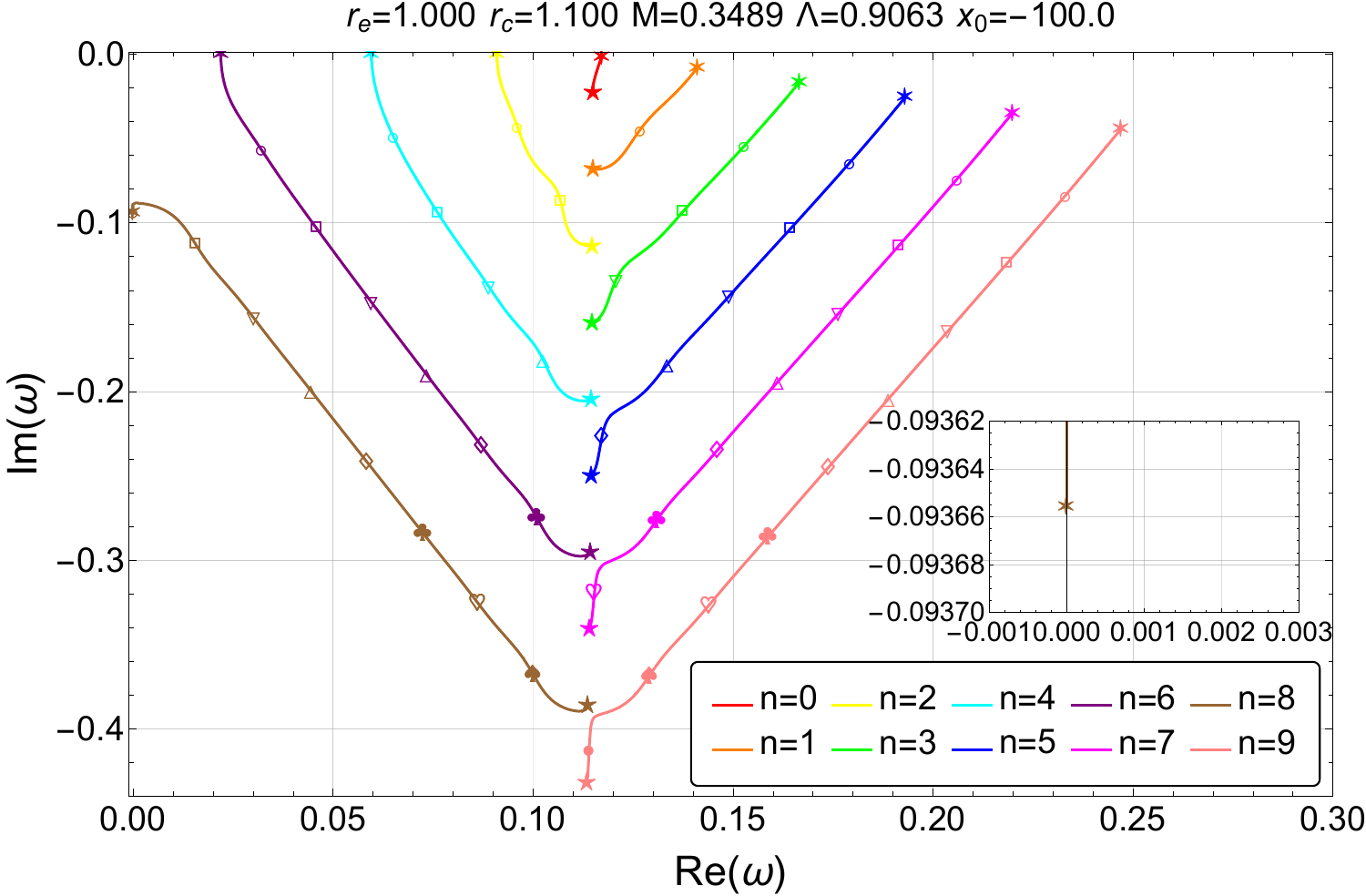}
    \caption{For the scalar perturbation $s=0$, the migrations of the QNM spectra with $\mathcal{K}\in[0,1]$ varying in the complex plane for the first $10$ modes, where the parameters are chosen $r_e=1$, $r_c=1.1$, and $x_0=-100$. Different color lines correspond to different modes.}
    \label{S_0_QNM_Spectra_near_extreme}
\end{figure}

Here we show a concrete result of QNMs for $s=0$ to clarify that the QNMs of other spins will have similar qualitative properties. In Fig. \ref{S_0_QNM_Spectra_near_extreme}, we show the migrations of the QNMs in the cases with $s=0$. From this figure, it can be seen that the qualitative migrations of QNMs are similar to the case of $s=2$.

\bibliography{reference}
\bibliographystyle{apsrev4-1}

\end{document}